\documentclass[10pt]{article}
\usepackage{enumerate}
\usepackage{caption}
\usepackage[pdftex]{graphicx}
\usepackage{epstopdf}
\usepackage{authblk}

\begin{document}

	\title{Elastic trapping of dislocation loops in cascades in ion-irradiated tungsten foils.}

	\author[1]{D R Mason\thanks{daniel.mason@ccfe.ac.uk}}
	\author[1,2]{X Yi}
	\author[3]{M A Kirk}
	\author[1]{S L Dudarev}
	\affil[1]{EURATOM/CCFE Fusion Association, Culham Science Centre, Abingdon, Oxfordshire OX14 3DB, United Kingdom}
	\affil[2]{Department of Materials, Oxford University, Oxford OX1 3PH, United Kingdom}
	\affil[3]{Materials Science Division, Argonne National Laboratory, Argonne, WI, USA}

	\maketitle

	\newcommand{\half}{\ensuremath{{^1\!/_2}}}

    \begin{abstract}
        Using \emph{in situ} transmission electron microscopy (TEM), we have observed nanometre scale dislocation loops formed when an
        ultra-high-purity tungsten foil is irradiated with a very low fluence of self-ions.
        Analysis of the TEM images has revealed the largest loops to be predominantly of prismatic $\half\langle 111\rangle $ type and of
        vacancy character.
        The formation of such dislocation loops is surprising since isolated loops are expected to be highly mobile, and should escape from
        the foil.
        In this work we show that the observed size and number density of loops can be explained by the fact that the loops are \emph{not}
        isolated - the loops formed in close proximity in the cascades interact with each other and with vacancy clusters, also formed in
        cascades, through long-range elastic fields, which prevent the escape of loops from the foil.
        We find that experimental observations are well reproduced by object Kinetic Monte Carlo simulations of evolution of cascades {\it
        only} if elastic interaction between the loops is taken into account. Our analysis highlights the profound effect of elastic
        interaction between defects on the microstructural evolution of irradiated materials.
    \end{abstract}

	\noindent{\it PAC61.72.Ff,61.80.Az,61.80.Hg,61.82.Bg}

	\noindent{\it Keywords \/
	in situ irradiations; dislocation loops; microstructural evolution}

    \section*{Introduction}

        The problem of microstructural evolution of tungsten under irradiation is attracting significant attention since tungsten has been
        chosen as plasma facing material for ITER divertor \cite{Bolt2002}, and also as a candidate material for the divertor of a fusion
        power plant \cite{Rieth2011,Rieth2013}.
        The advantages of tungsten as a divertor material are its high thermal conductivity, close to that of copper, the low sputtering yield,
        and the rapid decay of its activation following exposure to high energy neutrons \cite{Gilbert2011}.
        At the same time, tungsten is a brittle metal, particularly prone to brittle failure at grain boundaries.
        Experimental observations \cite{Bolt2002} suggest that the brittle to ductile transition temperature (BDTT) of tungsten increases as a
        result of exposure of tungsten to irradiation.

        Currently much less is known about the microstructure of irradiated tungsten than, say, iron alloys and steels, owing to the
        sparseness of experimental data, and particularly the small number of neutron irradiation studies \cite{Bolt2002,Thompson_pm1960}.
        It would be quite wrong to make assumptions about tungsten from our knowledge of iron alloys, as tungsten atoms are approximately
        three times heavier, have twice the cohesive energy, whereas the Debye temperature of tungsten is lower than that of iron although it
        is intended to operate at twice the temperature.
        Moreover, tungsten is non-magnetic, and so defect structures are different.

        In this paper we explore a particular aspect of microstructural evolution of irradiated tungsten related to the role played by elastic
        interaction between the defects formed in collision cascades.
        This question has received relatively little attention so far, and the potential significance of elastic interactions has largely been
        overlooked.
        New information, highlighting the importance of the above issue, became available recently through molecular dynamics (MD) simulations
        of very high energy collision cascade events \cite{Zarkadoula2013,Sand_EPL2013}, particularly through the analysis of size
        distributions of defects formed in such cascades \cite{Sand_EPL2013}.
        MD simulations showed that high-energy cascades often produce fairly large defects in close proximity to each other.
        Since the energy of elastic interaction is proportional to the product of defect relaxation volumes and varies as the inverse cube of
        the distance between the defects, we expect that microstructural evolution involving the formation of large defect clusters might be
        strongly affected by elastic interactions.

        We address the question by comparing experimental observations of very low dose self-ion implantation in ultra-high purity tungsten
        foil to computer simulations of cascade evolution in a foil.

        Our experiments, detailed in section \ref{experiment}, show that irradiation produces visible vacancy-type dislocation loops,
        predominantly prismatic loops with Burger's vector $\half \langle 111\rangle$, at the earliest stages of irradiation.
        That {\it vacancy}-type loops are produced is most likely an effect of foil surfaces \cite{Stoller_JNM2002,English_PM2010}, but the
        fact that prismatic loops still remain in the foil at experimental timescales (order seconds) in an ultra-pure material also requires
        explanation.

        Our experiments show a steady increase in loop size with temperature, starting already at room temperature, with loops
        doubling in area from 300 K to 773 K and then further quadrupling from 773 K to 1073 K, an average increase from 60 vacancies to 774
        per loop.
        Molecular dynamics simulations\cite{Sand_EPL2013} suggest only a handful of loops and a hundred isolated Frenkel pairs may be produced
        per cascade, and as our observations correspond to the limit of low fluence, inter-cascade interactions can be discounted.
        The enormous variation in loop diameter can not be accounted for solely by growth by accretion of vacancies, nor by coalescence of loops.

        The solution is found in the strong elastic interaction between loops.
        Langevin dynamics simulations and {\it in-situ} electron microscope observations of diffusion of dislocation loops have shown how
        elasticity correlates the motion of pairs of loops \cite{Dudarev_PRB2010}.
        In this paper we show how elastic interactions can dominate the evolution of the microstructure when loops can move along skew lines
        in close proximity to each other.
		The elastic trapping effect can retard the motion of prismatic dislocation loops for sufficient time for experimental observation.

        In section \ref{experiment} we describe the irradiation experiments and the TEM analysis of the defects formed.
        In section \ref{interactionEnergy} we detail the elastic interaction between defects, and show how this leads to a simple rule for the
        trapping of large dislocation loops.
        In section \ref{okmc} we sketch out the object Kinetic Monte Carlo simulations we have performed.
        Results from Monte Carlo simulations are shown and discussed in section \ref{results}.
        We show that elastic interactions are sufficient to trap loops in the vicinity of the original cascade volume for long enough to be
        seen experimentally.
        We find that at lower temperatures the largest vacancy loops generated in a cascade attract interstitial loops and shrink.
        At higher temperatures the largest loops are the most deeply trapped, and so are the only ones which survive to experimental times.
        This accounts for the observation of prismatic loops in an essentially pure metal, and their size and number variation with
        temperature.

	\section{Experimental procedure}
		\label{experiment}

		Tungsten sheets (Plansee Ultra-High Purity-W (UHP-W), $>$ 99.9999 wt\% pure, ~150 $\mu$m thick) were first cut into 3mm discs,
heat-treated in vacuum at 1673 K for 20 hours to allow thorough removal of dislocations and finally electropolished to reach electron
transparency thickness.
		The irradiations were performed at Argonne National Laboratory, using the IVEM-Tandem facility.
		150 keV W$^+$ ions were delivered through a tandem accelerator, incorporated with a 300 kV TEM.
		This allowed direct observation of the formation and evolution of defects.
		A double-tilt heating stage ensures that the diffraction conditions suitable for defect sizing and nature analysis could be set-up
precisely while the specimen is being heated.
		The temperature measurements were at 5 K accuracy.
		A Faraday cup installed at the entrance of the TEM column measured the dose.
		Note that the direction of the ion beam is 30$^\circ$ away from that of the electron beam due to engineering limitations.
		In practice, the TEM foil was tilted $\sim$15$^\circ$ to achieve a high incidence angle (75$^\circ$) for the ion beam, while not
sacrificing the microscopy quality.

		$(001)$ foils of UHP-W were irradiated up to $\sim$ $10^{16}$ W$^+$/m$^2$ at 300 K, 573 K, 773 K, 873 K and 1073 K respectively.
		The rate of dose was kept constant at $\sim$ $6.25\times10^{14}$ W$^+$/m$^2$/s.
		Based on an estimate of cascade radius of 5 nm (see Appendix \ref{initialDistribution}), we may consider that the defects observed in
this work were nucleated within distinct cascades.
		Detailed evidence for the non-interference of cascades at this dose is given elsewhere\cite{Yi_thesis_2013,Yi_PM_2013}.

		We have performed SRIM calculations\cite{SRIM} to compute the damage profile.
		The displacement threshold energy for a given PKA direction in a cubic crystal can be computed from experimental measurements using
the interpolation formula of Jan and Seeger\cite{Jan_PSS1963},
			\begin{eqnarray}
				E_d\left( \theta,\phi \right) &=& \left( 2 E_{d,110} - E_{d,100} \right) \nonumber \\
    &+& 2\left( E_{d,100} - E_{d,110} \right) \left(
    \sin^4\theta\left( \cos^4\phi+\sin^4\phi \right) + \cos^4\theta \right) \nonumber	\\
				&+& 9 \left( E_{d,100} - 4 E_{d,110} + 3 E_{d,111} \right) \sin^4\theta\cos^2\theta\sin^2\phi\cos^2\phi.
			\end{eqnarray}
		Integrating over all angles gives a single figure for the displacement threshold energy averaged over the directions of the incident beam,
			\begin{equation}
				E_d = \left( 2 E_{d,110} - E_{d,100} \right)\nonumber \\
 + \frac{6}{5} \left( E_{d,100} - E_{d,110} \right) + \frac{9}{105} \left(
E_{d,100} - 4 E_{d,110} + 3 E_{d,111} \right).
			\end{equation}
		For tungsten we use the experimental values of Maury \emph{et al.}\cite{Maury_RadEff1978}, to give $E_d=55.3$eV.
		Values of $E_d$ from $45$eV to $61$eV have been computed in molecular dynamics simulations using different empirical
potentials\cite{Fikar_JNM2009}.
		A threshold displacement energy for tungsten of $60eV$ has recently been shown to be the best fit to match low-energy Binary Collision Approximation (BCA) cascade
results to MD simulations\cite{Bukonte_NIMB2013}.

		The SRIM damage profile for 150 keV W$^+$ ions peaks at 10nm and extends up to 30 nm in depth, which is about half the thickness of a
typical thin region in UHP-W, deduced from an electron energy loss spectrum (EELS)\cite{Yi_thesis_2013,Egerton1996}.
		Our SRIM calculations suggest the experimental dose corresponds to 0.01 dpa.

	\subsection*{Experimental results}

\begin{figure}
			\centering
			\begin{minipage}{0.3 \textwidth}
			  	\centering
			  	\includegraphics[width=.9\linewidth]{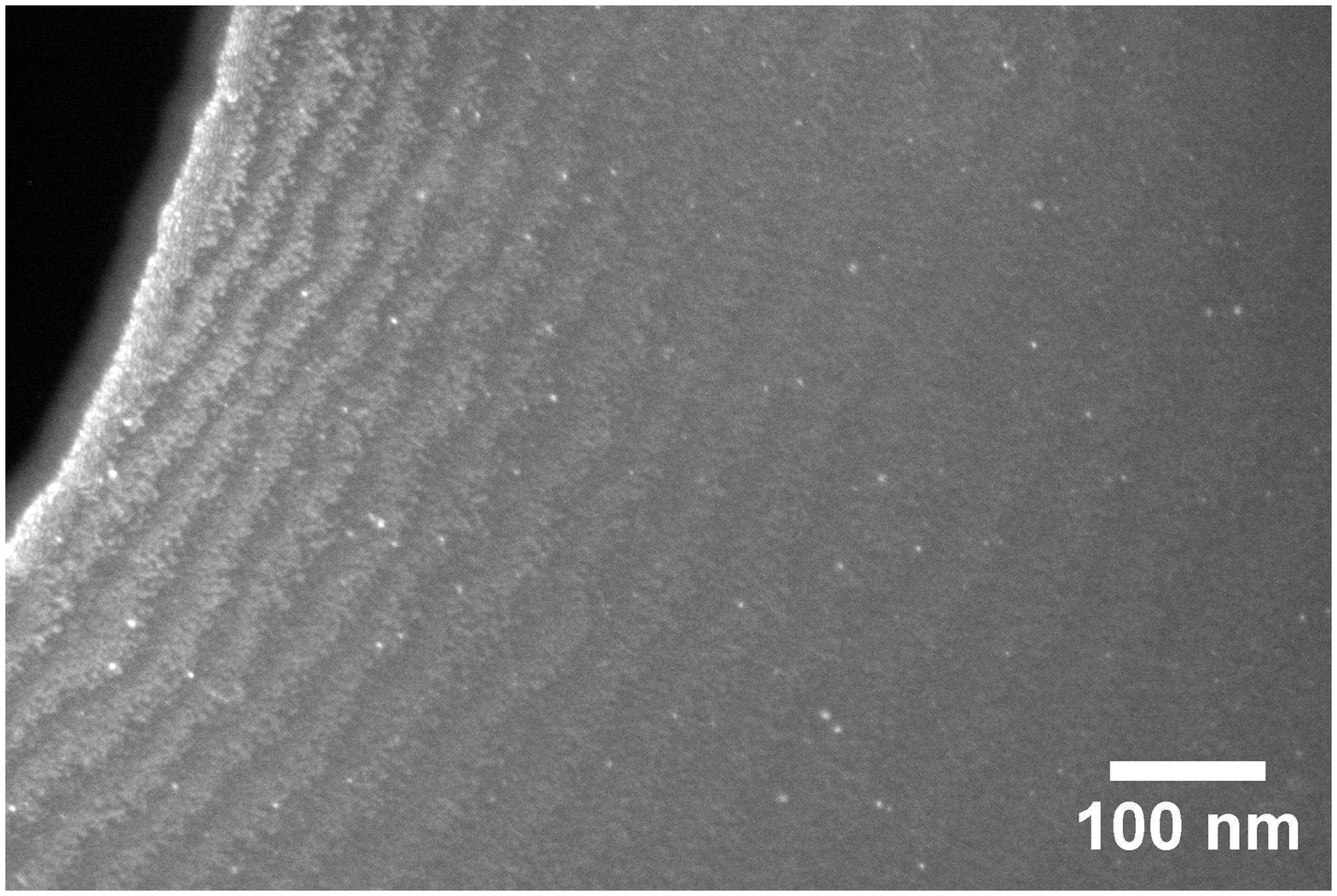}
			  	(a) 300K
			\end{minipage}
			\begin{minipage}{0.3 \textwidth}
			  	\centering
			  	\includegraphics[width=.9\linewidth]{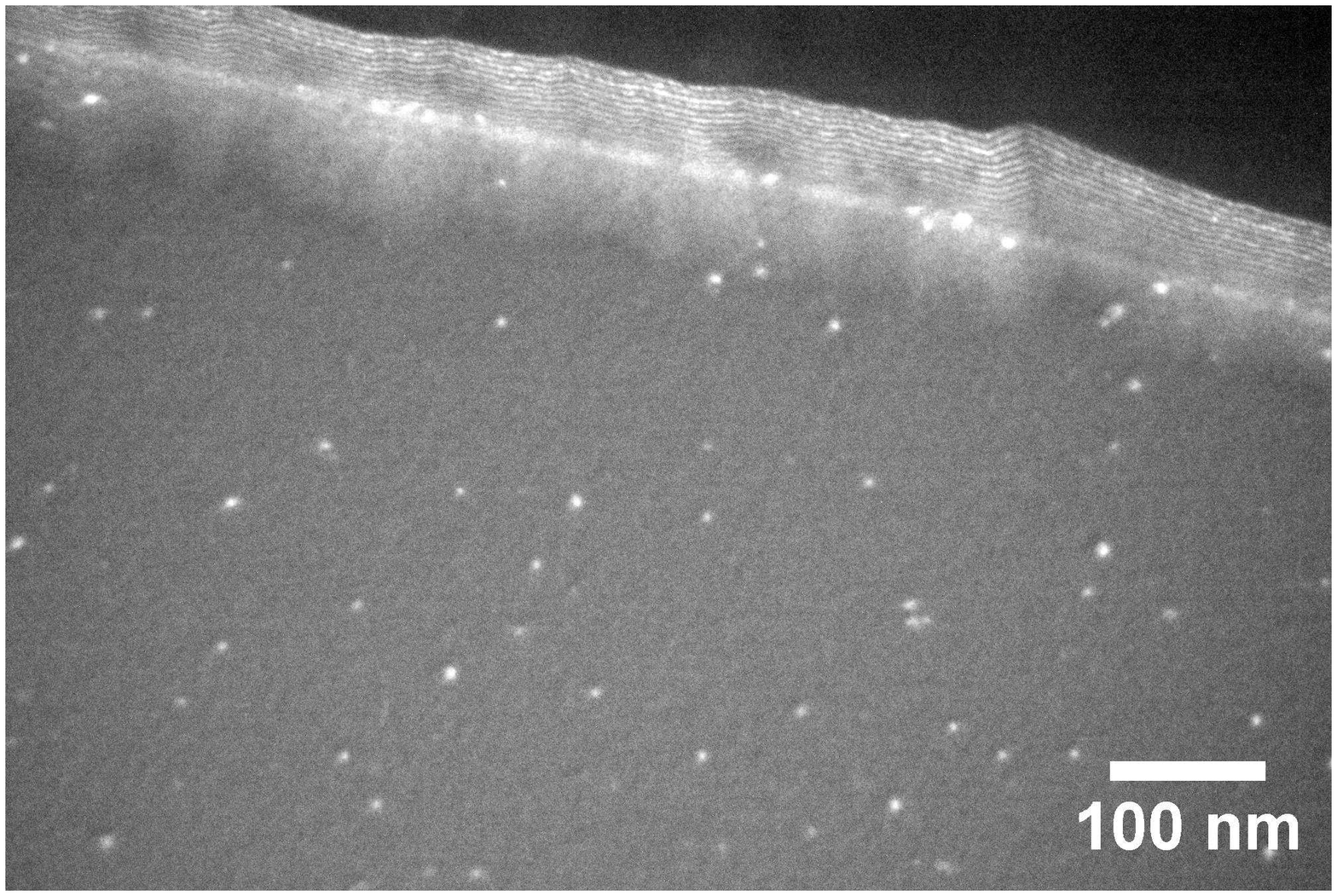}
			  	(b) 773K
			\end{minipage}
			\begin{minipage}{0.3 \textwidth}
			  	\centering
			  	\includegraphics[width=.9\linewidth]{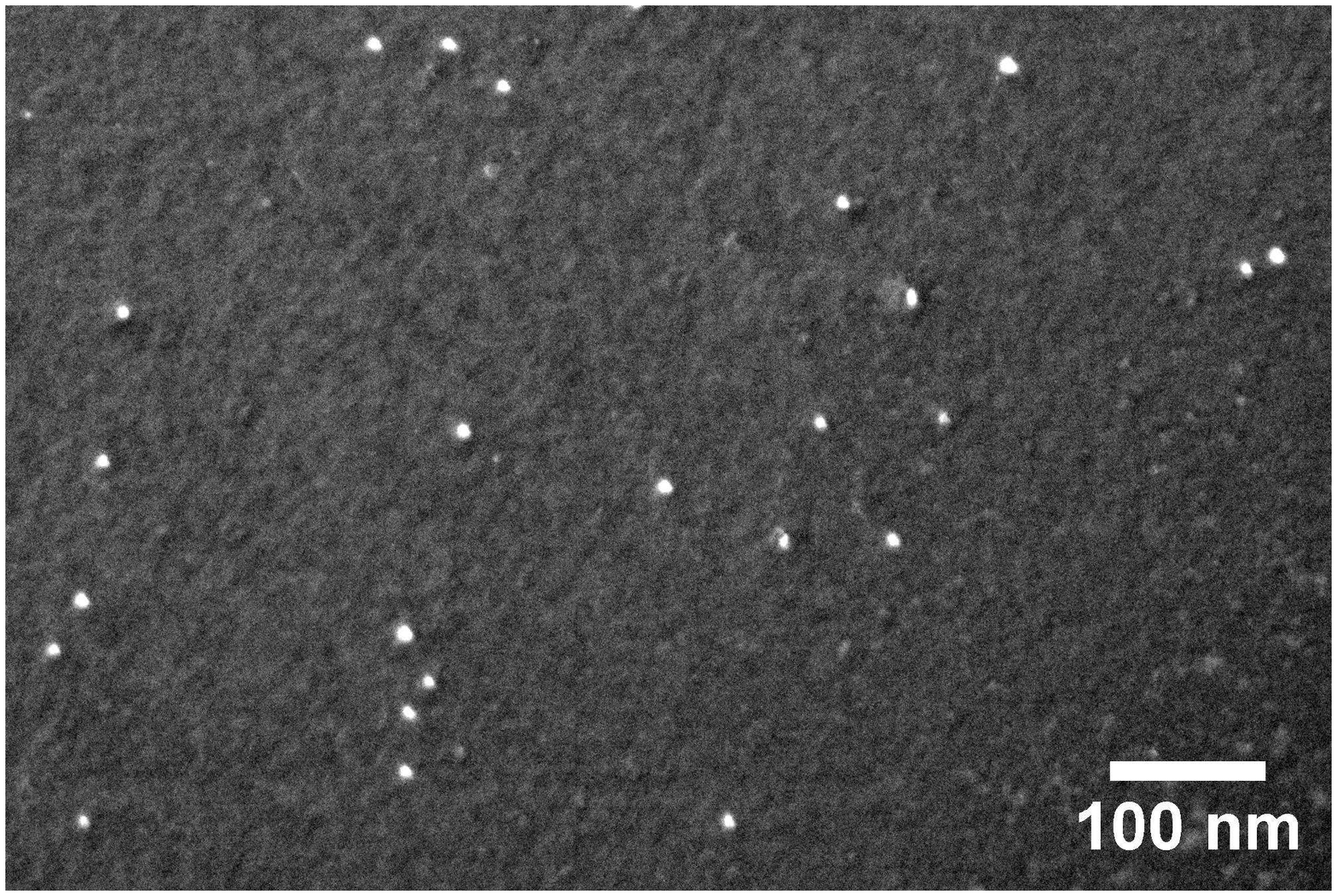}
			  	(c) 1073K
			\end{minipage}
		\caption{
		Damage microstructures imaged under weak-beam dark-field ({\bf g} = 200, 3-4{\bf g}) diffraction conditions in 0.01 dpa irradiated
UHP-W with differing irradiation temperatures.}
		\label{damage_micrographs}
		\end{figure}

		TEM micrographs recorded under weak-beam dark-field conditions ({\bf g} = 200, 3-4{\bf g}) were adopted for analyses of defect
population and size measurements.
		Figure \ref{damage_micrographs} shows representative areas of damage microstructure in tungsten at 300 K, 773 K and 1073 K
respectively.
		Strong temperature dependence can be found in terms of the defect yield (the ratio between the number of visible defects per unit area
and the ion dose) \cite{Jenkins1993} and the size of the defects.

\begin{figure}
			\centering
			\begin{minipage}{.50\textwidth}
			  	\centering
			  	\includegraphics[width=.9\linewidth]{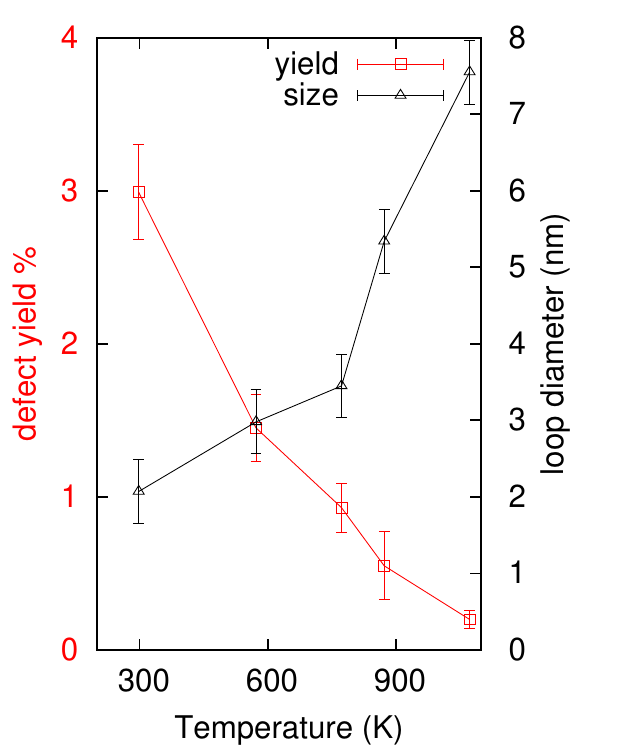}
			\end{minipage}%
			\begin{minipage}{.50\textwidth}
			  	\centering
			  	\includegraphics[width=.9\linewidth]{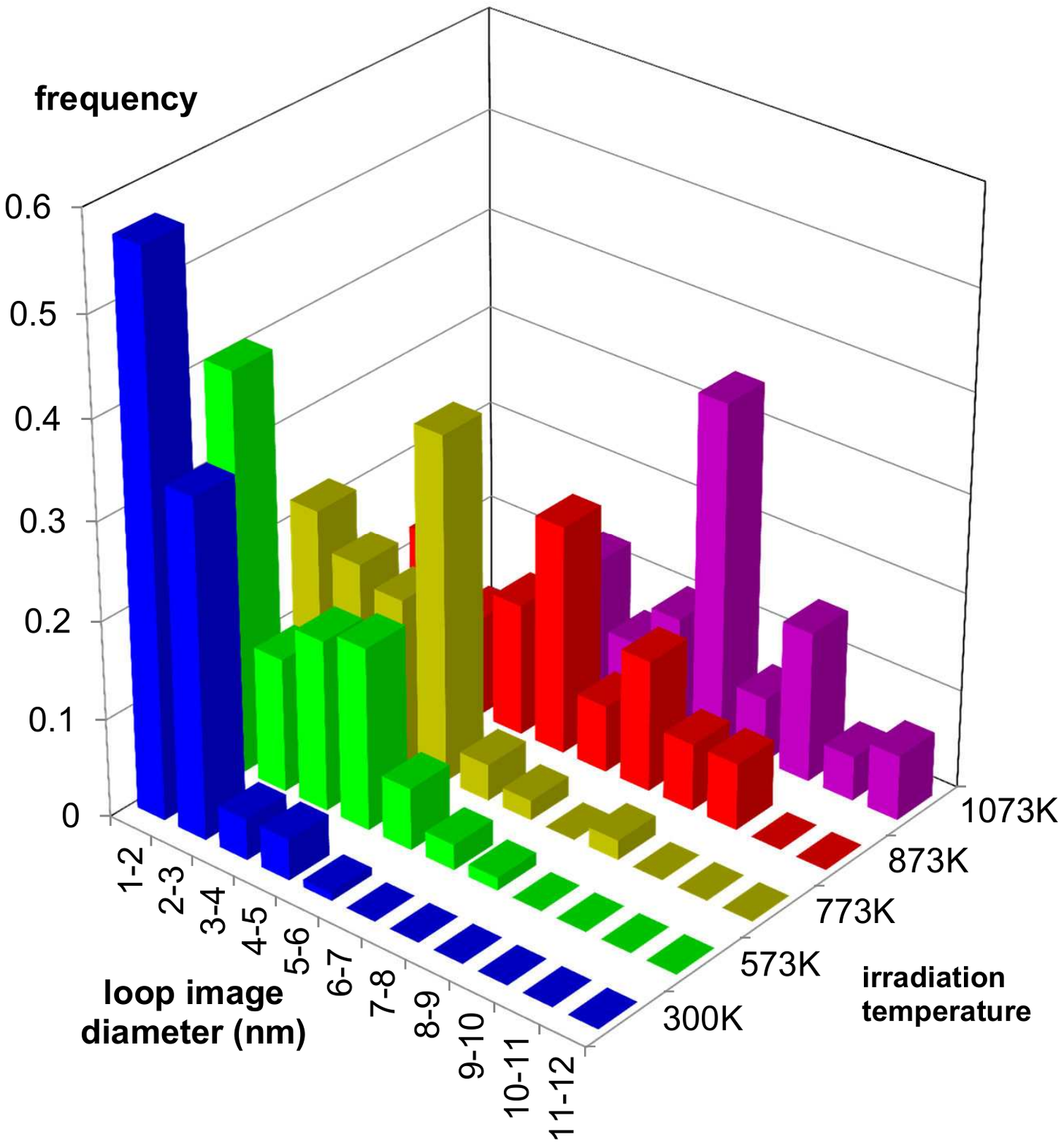}
			\end{minipage}
		\caption{
		Left: Defect yield ( number of loops seen per incident ion ) and average image diameter. Lines are to guide the eye.
		Right: Histogram of image sizes at different irradiation temperatures.}
		\label{expt_overview}
		\end{figure}

\begin{figure}
			\centering
			\begin{minipage}{1.\textwidth}
			  	\centering
			  	(a)
			  	\includegraphics[width=.75\linewidth]{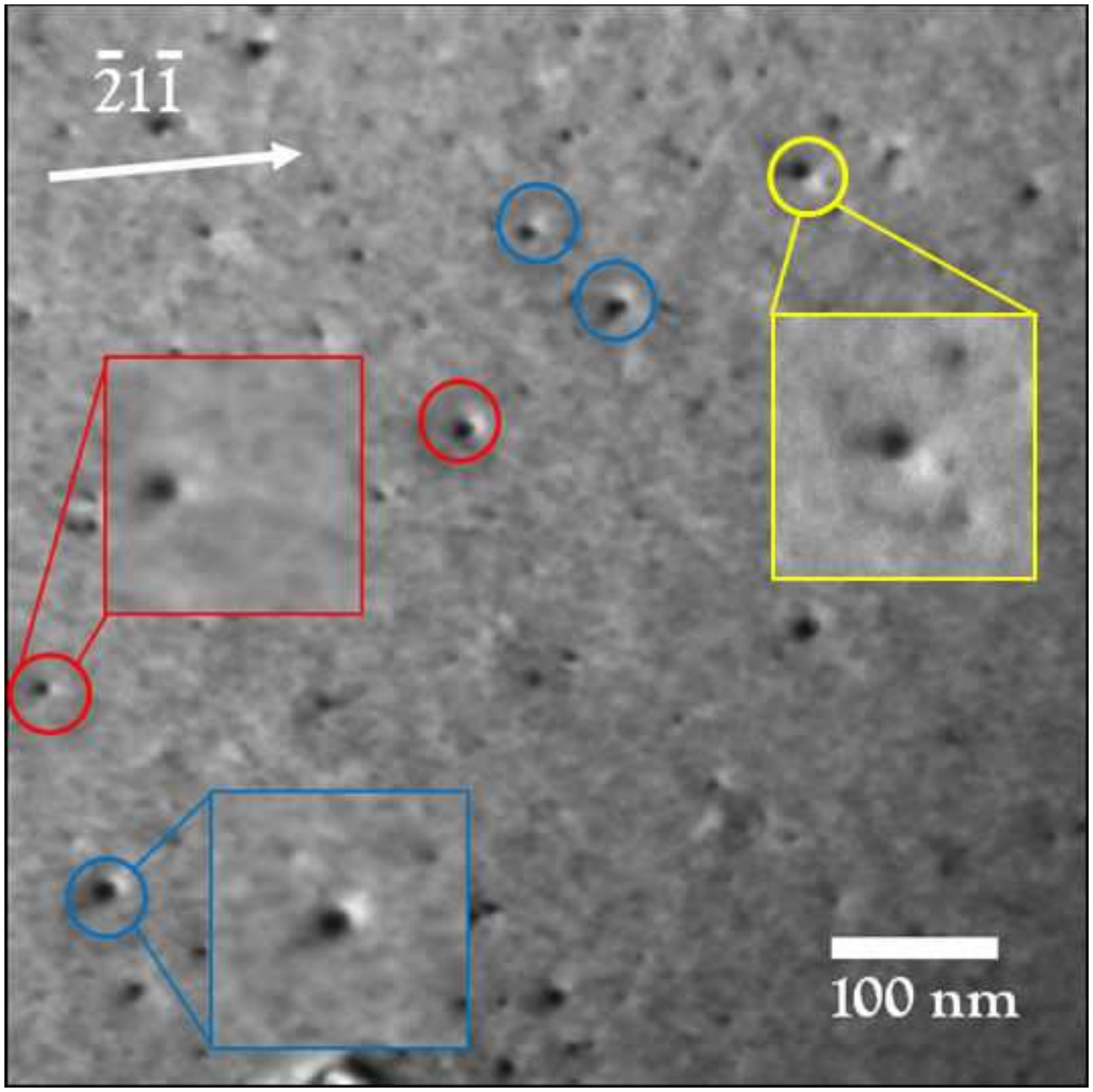}
			\end{minipage}%
			\newline
			\begin{minipage}{.3\textwidth}
			  	\centering
			  	\includegraphics[width=0.9\linewidth]{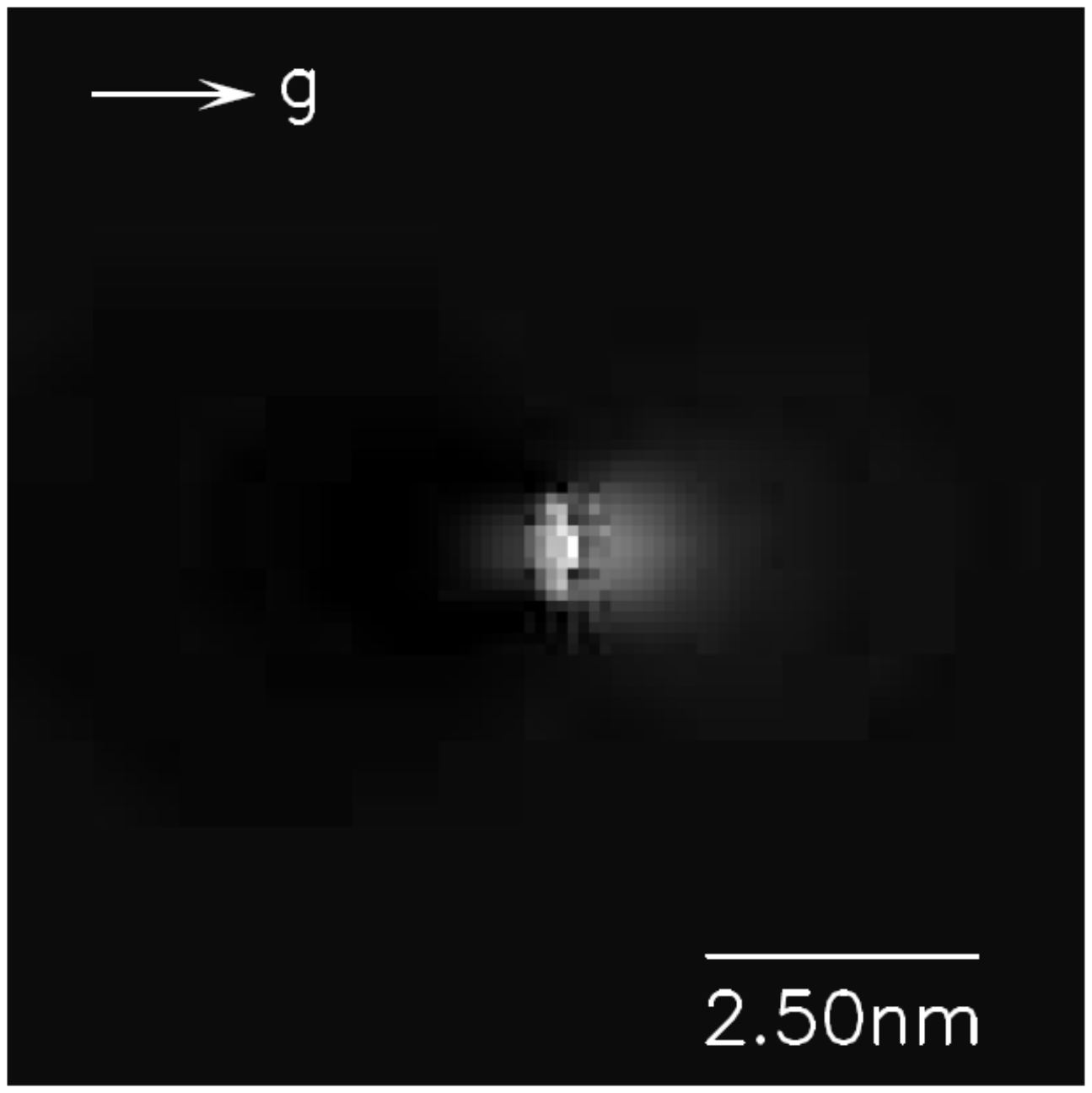}
			  	(b) sheared $\half\langle111\rangle$ vacancy loop (red).
			\end{minipage}
			\begin{minipage}{.3\textwidth}
			  	\centering
			  	\includegraphics[width=0.9\linewidth]{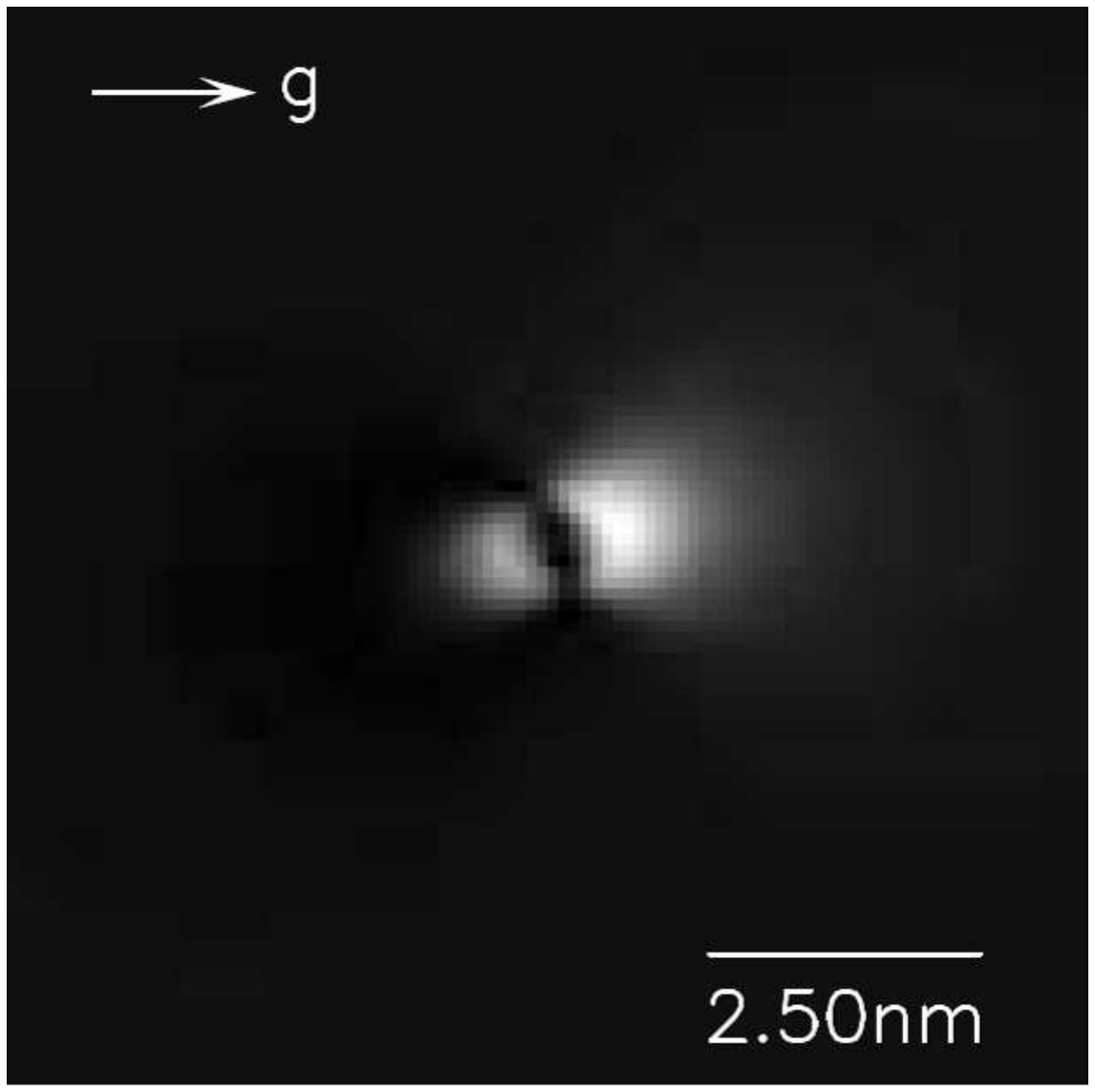}
			  	(c) pure edge $\half\langle111\rangle$ vacancy loops (blue).
			\end{minipage}
			\begin{minipage}{.3\textwidth}
			  	\centering
			  	\includegraphics[width=0.9\linewidth]{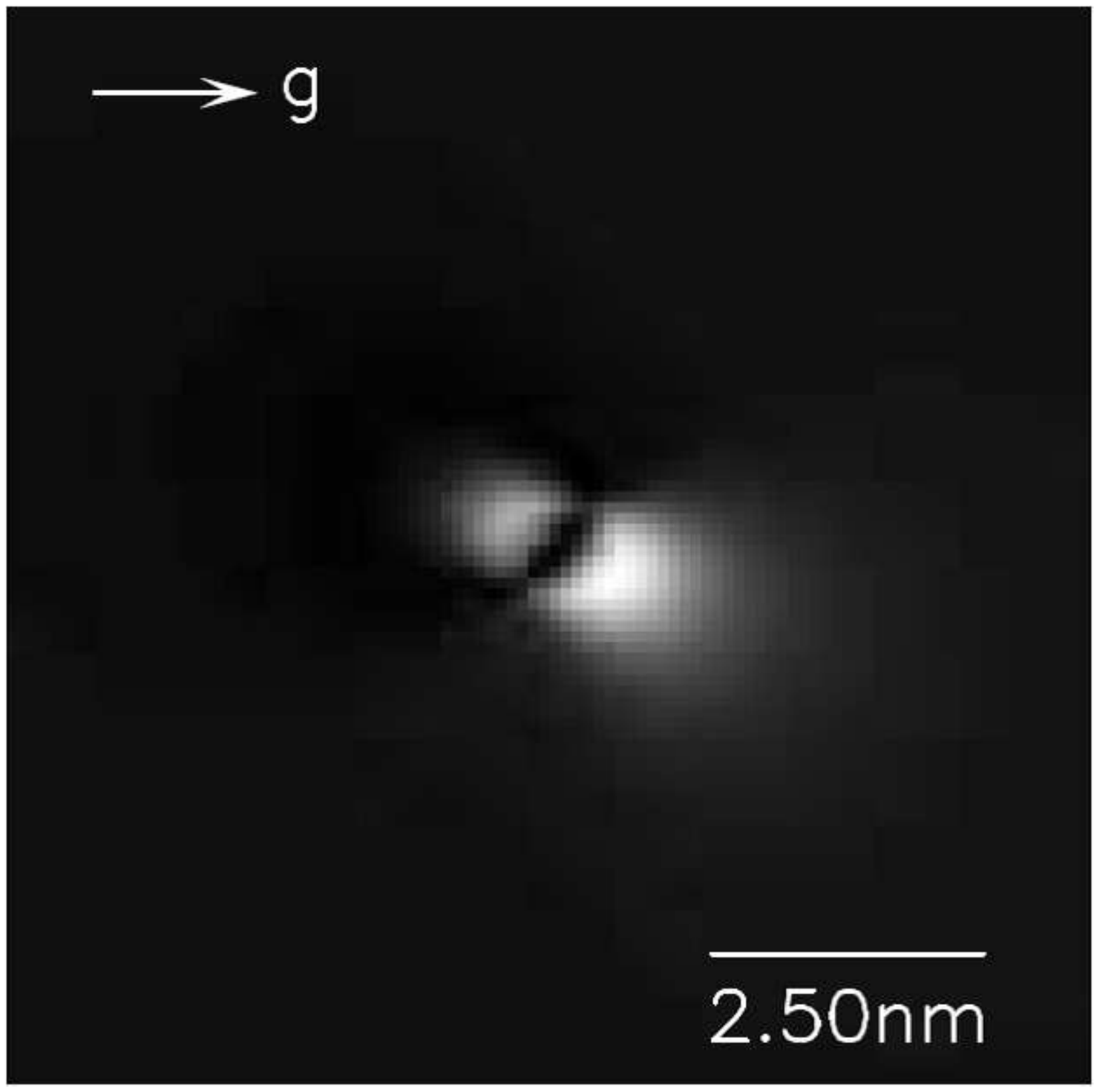}
			  	(d) $\langle100\rangle$ vacancy loop (yellow).
			\end{minipage}
		\caption{
		Comparison of black-white contrasts of dislocation loops obtained from irradiation experiment at 300 K and results of image
simulations using the TEMACI code\cite{Zhou_TEMACI_2005,Zhou_PRSA2005}.
		(a) TEM micrograph obtained from two-beam dynamical dark field condition, ({\bf g} = $\bar{2}1\bar{1}$).
		(b)-(d) Image simulations.
		}
		\label{LoopNatureIdentification}
		\end{figure}

		The efficiency of defect production through ion bombardments was low for all temperatures investigated; the highest being 2.99\% at
300 K.
		With the increase of irradiation temperature, defect yield dropped steadily, down to 0.2\% at 1073 K.
		A quantitative analysis of the temperature dependence can be found in figure \ref{expt_overview}.
		In contrast to defect yield, the average size of visible defects increased with the increasing temperature.
		The size of individual defects is shown in the plot of defect size distributions (Figure \ref{expt_overview}).
		Between 300 K and 573 K, defects with 1-2 nm diameters dominated in the microstructure; at 773 K, the largest fraction of defects
shifted to the 4-5 nm category.
		At 1073 K, 7-8 nm defects prevailed over other size categories.
		The geometry and nature of the observed defects were studied using the black-white diffraction contrast
technique\cite{Eyre1977_I,Eyre1977_II,Zhou_TEMACI_2005,Zhou_PRSA2005}.
		In figure \ref{LoopNatureIdentification}, we demonstrate the presence of various geometries of identified vacancy loops, i.e.
pure-edge and sheared $\half\langle111\rangle$ loops and $\langle100\rangle$ loops in tungsten irradiated at 300 K.
		A good match of diffraction contrast was found between experiment and image simulations.

		We also note from \emph{in situ} TEM observation of the irradiation that the loops generally appeared at maximum
brightness in a single frame (1/15s), although at higher temperatures a small minority of large loops are seen to brighten over a few frames.
Loops appeared immediately when the ion beam was switched on.
This is not consistent with a model which predicts a slow Ostwald ripening of the loops, nor relies on inter-cascade interaction.

    \section{Elastic interaction energy of objects}
    	\label{interactionEnergy}

		As tungsten is (almost) elastically isotropic\cite{Lowrie_JAP1967}, it is convenient for us to work with isotropic expressions for
elastic energy.
	    The isotropic elastic interaction energy between a pair of prismatic loops separated by a distance $R$ much greater than the loop
radii is
	        \begin{equation}
	            \label{Isotropic_elastic_loop_loop}
	            U_{12} = \frac{ \mu b_1 b_2 A_1 A_2 }{ 4 \pi R^3 (1-\sigma) } \left\{ \begin{array}{l} \displaystyle
	                        6(1-\sigma) \left[ (\hat{n}_1 \times \hat{n}_2)\cdot \hat{R} \right]^2
	                        - 2(1-\sigma)(\hat{n}_1 \times \hat{n}_2)^2 \\ [1ex]
	                         - 3 + 2(\hat{n}_1 \cdot \hat{n}_2)^2   \\ [1ex]
	                         + 3 (\hat{R} \cdot \hat{n}_1)^2  + 3 (\hat{R} \cdot \hat{n}_2)^2
	                         + 15 (\hat{R} \cdot \hat{n}_1)^2 (\hat{R} \cdot \hat{n}_2)^2   \\ [1ex]
	                         - 12 (\hat{R} \cdot \hat{n}_1) (\hat{R} \cdot \hat{n}_2) (\hat{n}_1 \cdot \hat{n}_2)
	                    \end{array} \right\},
	        \end{equation}
	        where loop $k$ has Burger's vector $b_k \hat{n}_k$ and area $A_k$. $\hat{R}$ is a unit vector along the loop separation direction,
$\mu$ is the shear modulus (161GPa in tungsten) and $\sigma$ is Poisson's ratio (0.28 in tungsten).
	        For practical usage in simulations, we regularise this formula at small separations by replacing $R^3$ in the denominator with
$R^3 + (\rho_1 +\rho_2)^3$.
	        Comparison to atomistic calculations using an empirical potential\cite{mason_2014} suggests this regularisation is good for the
interaction between vacancy and interstitial loops, but for like pairs of loops overestimates the repulsive energy when the loops come close
together.
	        As repulsive configurations will not persist we do not expect the dynamics to be affected significantly.
	    The elastic interaction between loop 1 and a vacancy cluster 2 is\cite{Bastecka_JPhysB1964}
	        \begin{equation}
	            \label{Isotropic_elastic_loop_cluster}
	            U_{12} = \frac{ \Delta V }{3 \pi} \left( \frac{ 1 + \sigma }{ 1 - \sigma } \right) \frac{ b \mu }{ \sqrt{ z^2 + (\rho + r)^2 }
}\left\{ \begin{array}{l} \displaystyle
	                        \frac{ \rho^2 - r^2 - z^2 }{ (\rho - r)^2 + z^2 } \mathrm{E} \left( \sqrt{ \frac{ 4 r \rho }{ z^2 + (\rho + r)^2 }
} \right)    \\ \displaystyle
	                            + \mathrm{K} \left( \sqrt{ \frac{ 4 r \rho }{ z^2 + (\rho + r)^2 } } \right)
	                    \end{array} \right\},
	        \end{equation}
	    where $z = \vec{R} \cdot \hat{n}_1$, $r^2 = R^2 - z^2$ and $\Delta V$ is the relaxation volume around the vacancy cluster.
	    $\mathrm{K}(k)$ is the elliptic integral of the first kind and $\mathrm{E}(k)$ the elliptic integral of the second kind.

\begin{figure}[h!t!b]
	           \begin {center}
	               \includegraphics [width=5.0in] {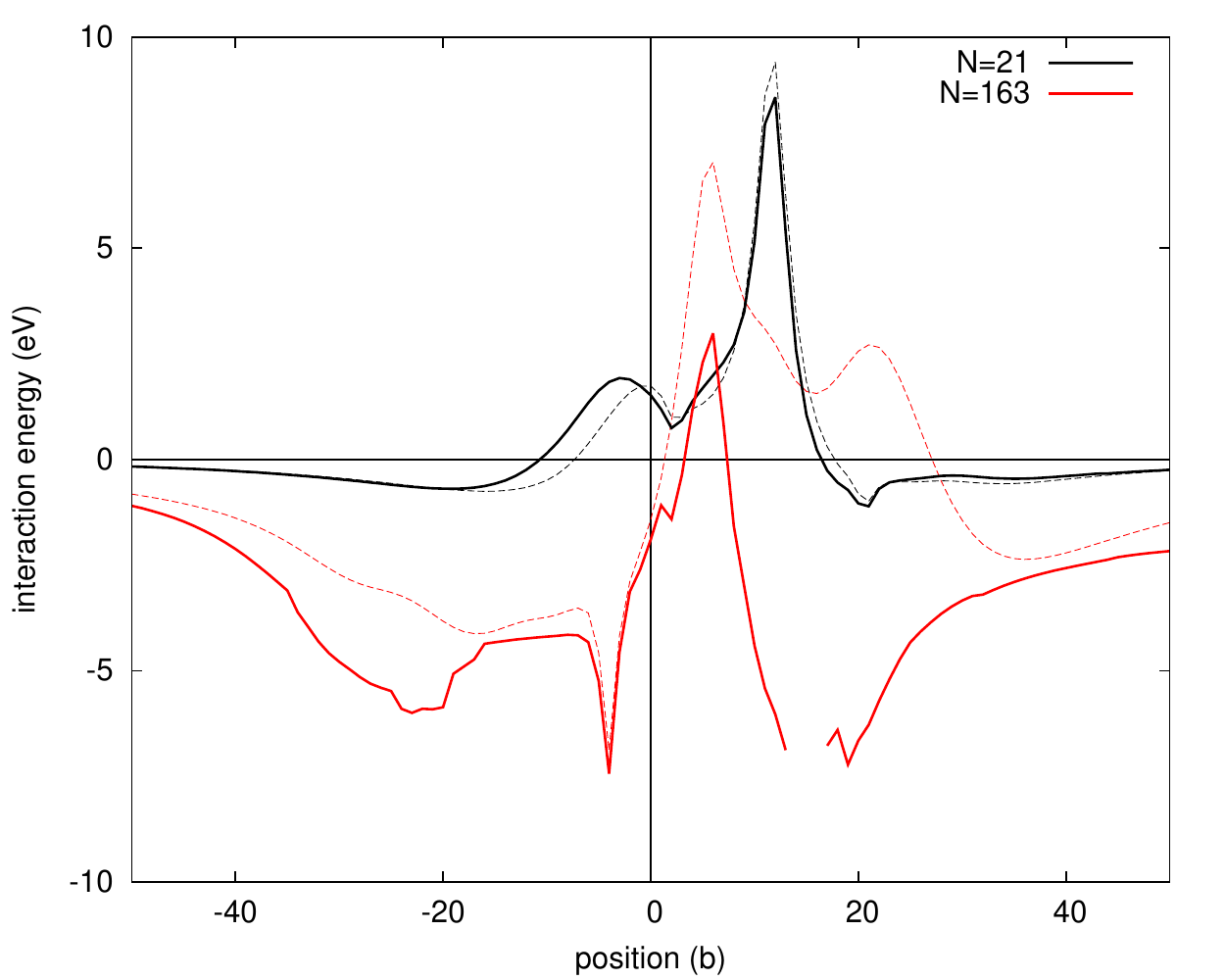}
	           \end{center}
	           \caption {
	           Typical potential energy landscape for a small and a large vacancy loop.
	           Note that the saddle points between discrete positions of the loop are not shown, and would contribute additional barriers to motion across this landscape.
	           A loop is taken from -infinity to +infinity and the total elastic energy of the configuration computed.
	           Regions where the loop would collide with another object are not drawn.
	           Dashed lines: other loops held fixed.
	           Solid lines: other loops allowed to move to a minimum energy configuration.
		       The curves are normalised to zero with the loop removed from the system.
	           	}
	           \label {drag}
	       \end {figure}

        The elastic interaction produces impediments to the free diffusion of loops in the forms of traps and barriers.
        The potential energy landscape seen by a typical loop dragged from -infinity to +infinity in bulk material is plotted in figure \ref{drag}.
        The loops were drawn from randomly generated configurations drawn using the procedure described in
        Appendix \ref{initialDistribution}.
        The landscape is computed with the assumption that all other loops are held fixed ( representing the highest barriers and shallowest
        traps );
        and again with other loops relaxed to the minimum energy configuration ( typically giving lower barriers and deeper traps. )
        Potential energy landscapes for a small vacancy loop containing 21 vacancies and a larger vacancy loop containing 163 vacancies are
        shown.
        Note that with one loop fixed there is no requirement for the curve at position 0 to be a minimum, and as a collective relaxation
        possible with the loop removed may not be possible with the loop fixed in one place, the relaxed curve is not required to lie below
        the unrelaxed curve.

\begin{figure}[h!t!b]
	           \begin {center}
	               \includegraphics [width=5.0in] {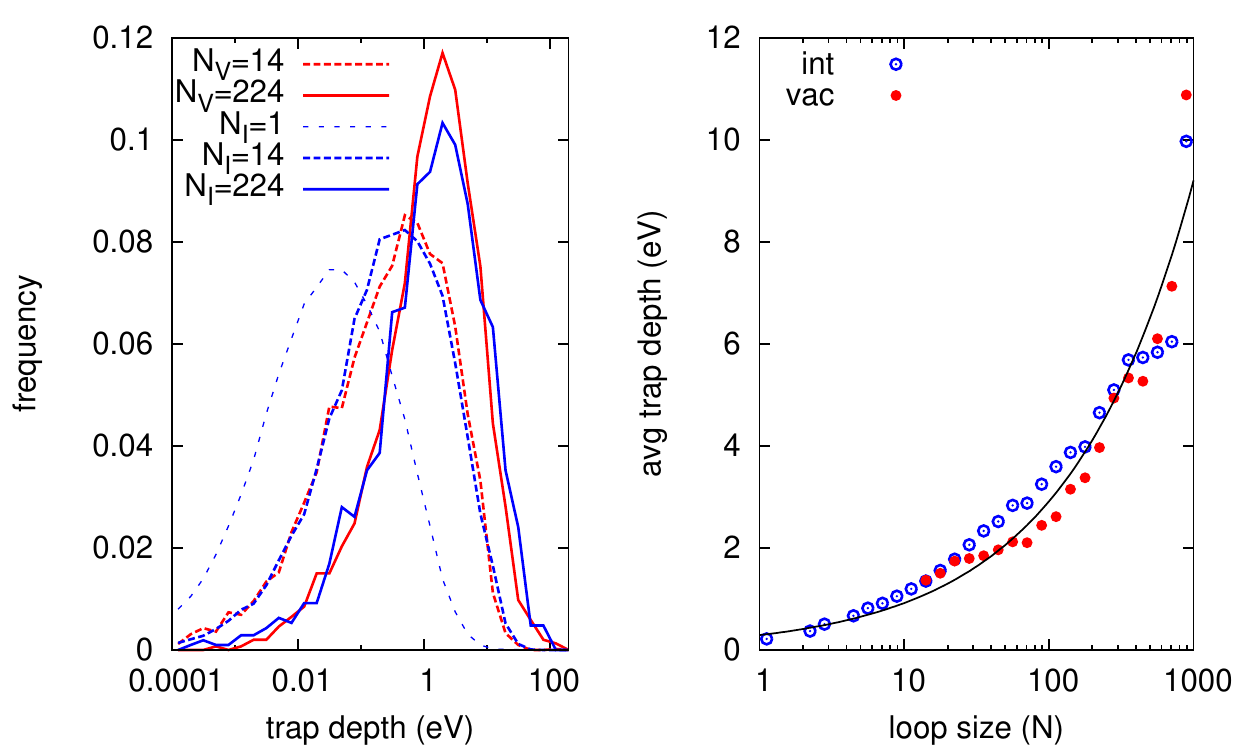}
	           \end{center}
	           \caption {
	           Left: Histograms of the elastic trap depth of loops at the initial configuration.
	           The bins are evenly spaced on a logarithmic scale, and the frequency represents the number of hits in each bin.
	           Right: mean depth of an elastic trap as a function of loop size.
	           The solid line is a fit to energy proportional to loop perimeter.
                }
	           \label {traps}
	       \end {figure}

		Figure \ref{drag} shows that the larger loop is seen to have much deeper elastic traps than the smaller loop in this instance.
		Statistics were generated for figure \ref{traps} from 24000 randomly generated starting conditions, this time in a foil of thickness
40nm.
		Trap depth is defined here as the difference in energy between a local minimum and the lower of the two maxima bounding it, and so is
a thermal barrier for evolution of the loop position rather than loop loss.
		The average trap depth increases with loop size as $\langle \Delta E\rangle  = (0.291 \pm 0.007) \sqrt{N}$ eV, suggesting the trap depth is
proportional to the length of the perimeter of the loop.
		The most attractive elastic energy condition occurs when loop perimeters pass close together.
		Interstitial loops generated on the outside of the cascade do not often pass close together, and in nearly 90\% of configurations zero or one vacancy loops are generated.
		Hence the most attractive energies typically occur when a vacancy loop passes an interstitial loop and the loop normals are skew.
		There is no significant difference between trap depths for vacancy- type and interstitial- type loops using our initial configurations.

		While this case is somewhat idealised, as it does not account for collective motion, figure \ref{traps} nevertheless demonstrates
there is ample elastic interaction energy between loops to cause a significant impact on their dynamic evolution.

    \section{Object Kinetic Monte Carlo}
    \label{okmc}

	    Our choice of simulation technique is constrained by the requirement that the simulations reach experimentally observable time-scales,
and that adequate statistics be generated.
	    While large molecular dynamics simulations can contain enough atoms to model a reasonable section of foil (a cube of side $\sim$40nm
is needed, containing 4M atoms), they are limited to modelling the first few nanoseconds at best and this would be an expensive calculation to
repeat.
	    Off-lattice atomistic kinetic Monte Carlo and accelerated molecular dynamics would, in all likelihood, struggle to reach the system
size needed, and would not necessarily outperform MD as the many mobile entities means a very large number of possible saddle points out of
each metastable atomic configuration.
	    We have therefore chosen to simulate the evolution of interacting defects using object Kinetic Monte Carlo (okMC).
	    okMC has been successfully used previously in many studies of radiation damage
recovery\cite{Voter_RadEff2007,Domain_JNM2004,Stoller_JNM2008,Becquart_JNM2009}.

\begin{figure}[h!t!b]
	        \begin {center}
	            \includegraphics [width=5.0in] {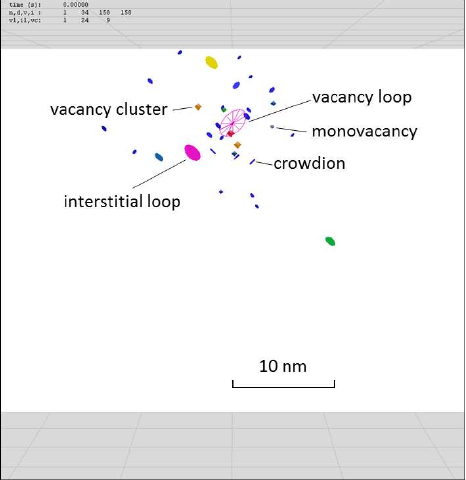}
	        \end{center}
	        \caption {
	        A typical cascade of 158 Frenkel pairs arranged into 34 defects at time t=0. The large vacancy loop in the centre contains 113
vacancies ( 2.8 nm diameter ).
             }
	        \label {Picture1}
	    \end {figure}

	    The success of okMC is dependent on the care with which the possible entities and their interactions are defined.
	    For our simulations four classes of object are defined: the vacancy cluster, the vacancy loop and the interstitial cluster and
interstitial loop.
	    These are idealised to perfect geometries, and collisions between them handled extremely simply by determining a capture length (see
table \ref{GeometryOKMC}).
	    It is hoped that these acknowledged weaknesses are mostly in the handling of the interaction of colliding objects, ie short time-scale
phenomena, and that the long-time scale diffusive motion is broadly correct.
	    The initial configurations are generated randomly according to the procedure in Appendix \ref{initialDistribution}; an illustrative
initial condition is shown in figure \ref{Picture1}.

		During the cooling phase of cascade impurities have little or no effect on the defect structures produced, as the timescale for recrystallisation is too short to permit any impurity segregation.
		As we are modelling an ultra-pure (99.9999\% pure) sample, it is reasonable to consider the cascade volume itself (5nm radius) as pure metal.
		Each cascade contains dislocation loops, plus clusters and point defects, all produced within some tens of nanometers of the surface and undecorated by impurities.
		Isolated prismatic loops would then move extremely quickly to the surface, with no time to pick up an impurity atmosphere.
		For these reasons it is quite reasonable to model the processes of loop generation and loss to surface as occurring in a perfectly pure material.

	    The monovacancy is permitted to move by nearest neighbour hopping. The activation barrier is taken to be 1.78eV, established by
DFT\cite{Nguyen-Manh_PRB2006} and experiment \cite{Rasch_PMA1980}.
	    The rate prefactor for monovacancy migration is taken as four times the Debye frequency $\nu_0$, to reproduce the experimental
self-diffusion coefficient\cite{Mundy_PRB1978}.
	    We have computed migration barriers and rate prefactors with an empirical tungsten potential\cite{mason_2014} for a vacancy in the
vicinity of a prismatic loop, and a cluster.
	    The barrier is seen to be hardly affected until the vacancy comes into the region of the dislocation core.
	    The rate prefactor is sensitive to position, showing the migration entropy $S_m$ ( defined by $\Gamma = \nu_0 \exp( S_m/k_B ) \exp(
-H_m/k_BT )$ ) varying by a few $k_B$ as the vacancy comes up to the larger defect.
	    Unfortunately it was not possible to extract any functional form to predict the rate prefactor as a function of position, so we leave
it as a constant.
	    Details of all energies and rates are given in Appendix \ref{parameterization}.

	    Vacancy clusters can undergo Brownian motion.
	  	Molecular dynamics of $N=4-13$ clusters at 1500K for 10ns using an empirical potential\cite{mason_2014} has failed to find any low
energy pathways for surface rearrangement.
	  	We have also looked at all surface rearrangements for small vacancy clusters and larger voids involving one atom moving to a nearest
neighbour position using a nudged elastic band method and the same potential.
	  	The minimum surface rearrangement barrier we find is comparable with the vacancy migration barrier.
	  	The probability of Brownian motion is proportional to the number of high energy surface atoms, which we take to be proportional to
surface area for $N\le 13$, and constant for $N\ge 13$.

	  	Clusters can absorb monovacancies or other clusters whose surfaces come within a recombination length of 1 unit cell. The larger
cluster absorbs the smaller.
	  	Vacancy clusters are also permitted to eject monovacancies to a position outside the recombination length in one of the eight
$\langle111\rangle$ directions, with a thermal barrier equal to the vacancy migration barrier.

	    The elastic interaction between monovacancies and vacancy clusters is very small beyond one unit cell
separation\cite{Ventelon_JNM2012}, and identically zero in isotropic elasticity,
	    so we treat pairs of clusters as non-interacting.

	    Isolated prismatic vacancy loops are highly mobile, comparable to interstitial loops\cite{Derlet_PRB2011}.
	    We have found vacancy loops of diameter $>$1nm to be metastable at low temperatures, with dislocation loops stable to rearrangement in
MD simulations over tens of nanoseconds at 1200K\cite{mason_2014}, in agreement with experiment but in contrast to work with an earlier
interatomic potential\cite{Gilbert_JPCM2008}.
	    MD also shows that the formation energy and mobility of loops with irregular edges is similar to that of circular loops.

	    We model all loops as $\half \langle 111\rangle $ Burger's vector prismatic circular loops. This simplification is to avoid the necessity of
introducing parameters for properties of $\langle 100\rangle $ Burger's vector loops, or loops with complex structures.
	    Loops may move in one dimension along their principal axis, and are also permitted to move perpendicular to their normal direction via a
movement of atoms in the dislocation core.
	    This latter movement is given a high thermal barrier, and moves the centre of position by $a_0\sqrt{3/4}/N$.
	    It is introduced to allow like-character loops elastically trapped close together to coalesce, as observed in experiment\cite{Arakawa_ActaMat2011}.

	    Loops interact with each other and with clusters through elastic fields using periodic boundaries in the x- and y- directions and the
minimum image convention.
	    In our model we neglect the image-stress interaction between a loop and the surface, as this is very small unless loops come
sufficiently close that they may be absorbed into the surface anyway.

	    Loops can absorb other smaller loops and vacancy clusters, and vacancy loops can eject monovacancies to a position just beyond its
perimeter.

	    Interstitial clusters can move like interstitial loops but are also permitted to rotate/change their Burger's vector.

		Any loop or cluster which intersects the surface is immediately removed.

	    A kinetic Monte Carlo code was used to evolve the objects through time according to these laws, using the standard process of choosing
the next event to occur with a probability proportional to the rate of that event.
	    Five modifications to the standard procedure were employed to speed up the evolution.
	    \begin{itemize}
	    	\item
	    	The mixed first- and second- order chain RTMC algorithm was used to eliminate short- lived flicker
events\cite{Mason_CPC2004}.
	    	\item
	    	A tree of visited states was constructed on-the-fly. Barriers between states were stored, and only when a new state is visited
do new barriers need calculating.
	    	These first two methods do not introduce any approximations.
	    	\item
	    	For the highest temperatures considered (900K and 1050K) only very large loops remain to t=1s, yet these are only generated
very infrequently in the starting configuration.
	    	As we know the probability of generating a large loop, we can properly weight starting configurations artificially seeded with
large loops to improve our statistics of rare events.
	    	For 900K and 1050K, we used 1000 randomly generated configurations plus a further 100 configurations seeded with a vacancy
loop containing more than 400 defects.
	    	\item
	    	Barriers between events involving object 1 were considered not to have changed if an event involving object 2 occurred,
provided there is a minimum separation of 10 lattice parameters between the perimeters of the objects.
	    	This sort of localisation approximation is standard in on-the-fly type kinetic Monte Carlo
simulation\cite{Mason_JPCM2004,Xu_PRB2011}.
	    	\item
	    	Where the elastic energy of \emph{all} loops is lower than $-10 k_B T$, escape from the traps becomes infrequent, although the
loops continue to jump back-and-forth within the trap.
	    	If this elastic energy criterion is met for 1000 consecutive steps, the loops are assumed to be essentially sessile \emph{for
the following step}.
	    	In place of their standard one-Burger's-vector move, a single jump removing them from the foil is introduced as a function of
the distance from loop to foil surface and the change in elastic energy to the surface.
	    	In practice, as these loop escapes are low-rate events, another infrequent event (such as a vacancy movement) generally occurs as the next step, moving the system on in time and helping to remove the trap.
	    \end{itemize}

      \section{Simulation Results}
      \label{results}

	    \begin{figure}[h!t!b]
	        \begin {center}
	            \includegraphics [width=5.0in] {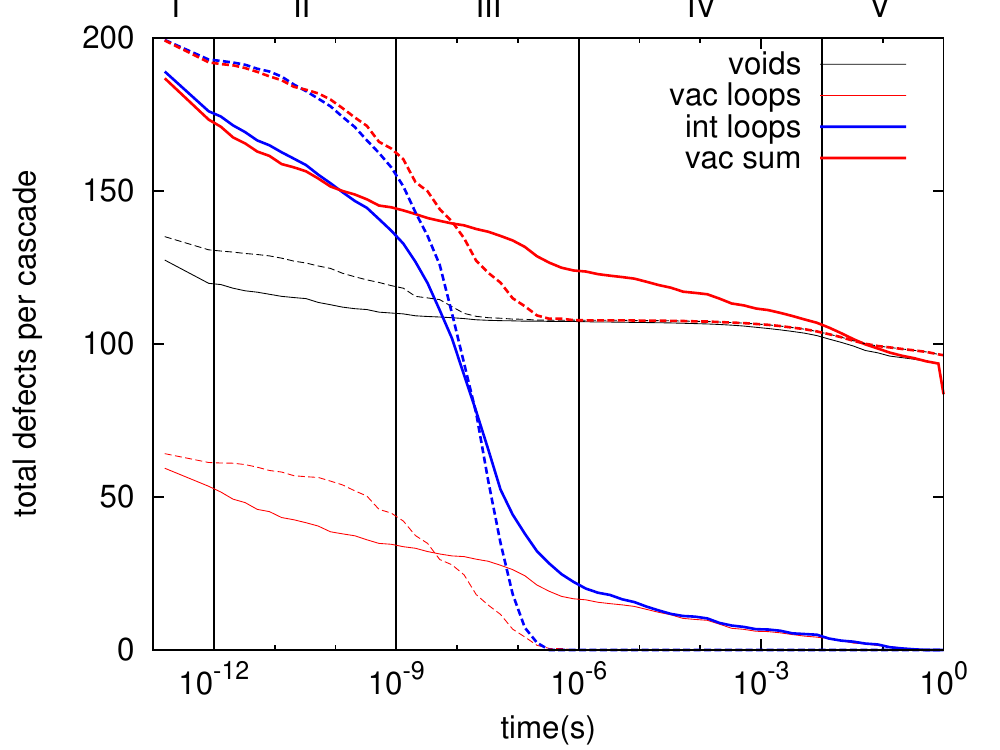}
	        \end{center}
	        \caption {
	        The average number of interstitials and vacancies in defects computed as a function of time for T=1050K.
	        Vacancies are divided into those in voids (including monovacancies), those in loops and their sum.
	        Results generated with and without elastic interactions are shown with solid and dashed lines respectively.
             }
	        \label {timeevolution2}
	    \end {figure}

		Figure \ref{timeevolution2}	shows the average number of interstitials and vacancies remaining after a cascade as a function of time
for T=1050K.
		Note that at time t=0 there is an average 200 Frenkel pairs, following the number found in MD simulations\cite{Sand_EPL2013}.
		Five stages of recovery can be seen. At different temperatures these same stages are observed, albeit at different times.
		\begin{enumerate}[I]
			\item
				In the first few picoseconds only crowdions are mobile, and are lost to vacancy sinks and the surface of the foil.
			\item
				From picoseconds to one nanosecond the loops are mobile, and intra-cascade recombination occurs.
			\item
				From one to one hundred nanoseconds loops ( both vacancy and interstitial ) are lost to the surface of the foil.
			\item
				From one hundred nanoseconds to ten milliseconds vacancy clusters are sessile. Some remaining loops not in deep elastic
traps continue to be lost.
			\item
				After ten milliseconds monovacancies, then larger vacancy clusters become mobile and are lost to the surface.
		\end{enumerate}

		Elastic interactions show a very pronounced effect on the recovery at all stages.
		In stages I and II the rate at which interstitial and vacancy loops generated in the same cascade recombine is significantly increased if they see each other
through long-range elastic interactions.
		Compared to the non-elastic calculation, almost twice as many interstitials have recombined by 0.1ns.
		In stage III all interstitial and vacancy loops are lost to the surface if there are no elastic interactions to trap them.
		The number of vacancies in voids in stages III,IV and V is similar whether elastic interactions are present or not.
		Importantly we note there is little transfer of vacancies from voids to loops in these latter stages- individual loops are not
significantly growing in size.
		In stage IV there are still an appreciable number of loops present, generated in the same cascade and bound together with elastic interactions, with a roughly equal
number of vacancy and interstitial loops.
		In stage V the gradual removal of monovacancies and shrinkage of voids reduces the elastic traps and more loops are lost.

	    \begin{figure}[h!t!b]
	        \begin {center}
	            \includegraphics [width=5.0in] {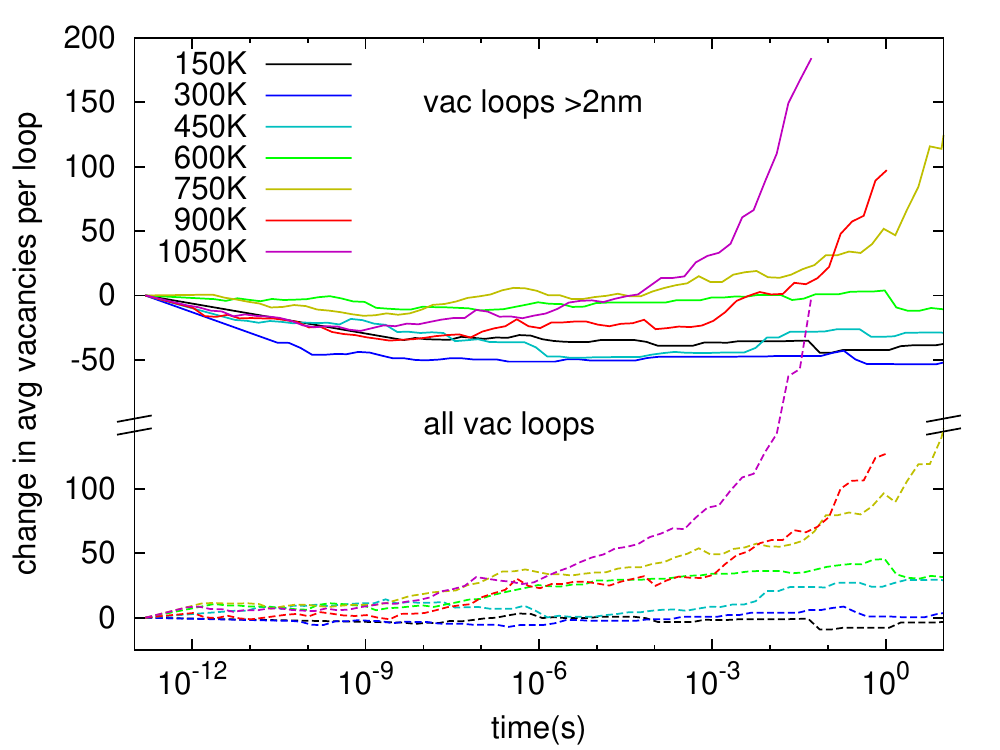}
	        \end{center}
	        \caption {
	        The change in average number of vacancies per loop computed as a function of time at different temperatures.
	        Upper plot : change in number of vacancies per large loop ( diameter $> 2$nm ). At time t=0 the average number is $217\pm 1$.
	        Lower plot : change in number of vacancies in all loops, visible or not. At time t=0 the average number is $109\pm 1$.
             }
	        \label {timeevolution}
	    \end {figure}

	    Figure \ref{timeevolution} shows the average number of vacancies per vacancy loop as a function of time and temperature.
	    When all loops are considered, there is a monotonic increase in size of loops, with the gradient increasing in stage IV.
	    This effect is due to the smallest loops being most mobile, and so escaping to the surface.
	    While there is some stochastic variation owing to the small number of independent samples, as a rule the simulations at higher
temperature can detrap increasingly larger loops, hence at higher temperature a larger average diameter is seen.

	    When only the large loops are considered there is a significant reduction in size at short times, and this reduction is most
pronounced at lower temperatures.
	    This effect is again elastic in origin, and corresponds to the largest vacancy loops attracting interstitials and
thus most rapidly shrinking.

	    \begin{figure}[h!t!b]
	        \begin {center}
	            \includegraphics [width=5.0in] {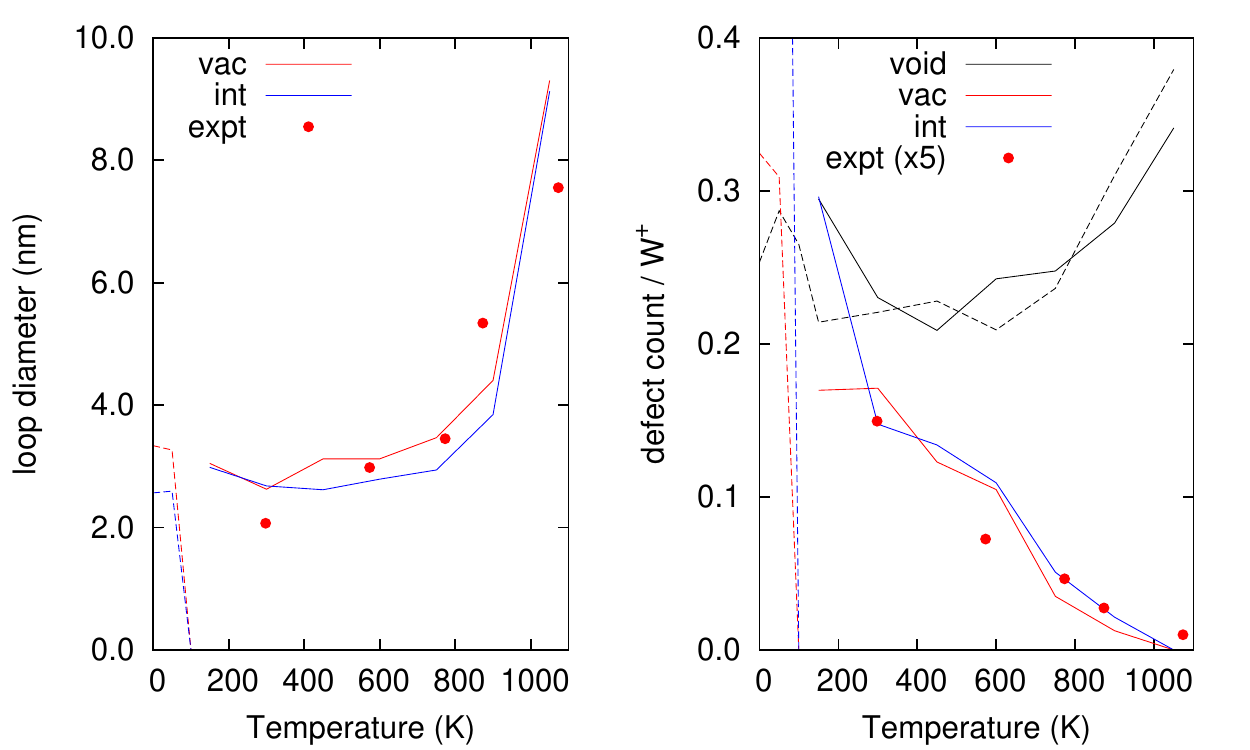}
	        \end{center}
	        \caption {
	        The average size of visible defects at time t=1s, and the number of visible (diameter $>1.5$nm) defects per cascade.
	        Solid lines: with elastic interaction included.
	        Dashed lines: without elastic interaction.
	        Points: experiment (vacancy loops).
	        The number of visible defects seen experimentally has been multiplied by 5 to fit on this scale- justification for this adjustment
of the absolute numbers is given in the text.
	        }
	        \label {visible_t=1s}
	    \end {figure}

	    Figure \ref{visible_t=1s} shows the size and number of (potentially) visible defects with diameter $>1.5$ nm, with a comparison to the
experimental data.
	    It is immediately seen that when elastic interactions are included simulations well reproduce experimental trends and
scales, without elastic interactions there are no remaining loops at observable times.

	    The number of visible vacancy loops in the simulations is approximately five times that seen in experiment, but the trend in relative
numbers is reproduced.
	    It is the trend and not the absolute numbers which is significant.
	    The absolute numbers of loops seen depend on a number of factors- most importantly on the relative probability of generating vacancy
loops versus voids produced in a cascade, but also on the minimum size of a loop deemed observable, the number of loops rendered invisible in
the TEM by meeting the $\mathbf{g}\cdot\mathbf{b}=0$ criterion, and on the fate of loops between the time where one second has elapsed and the
point of experimental observation.
	    These factors are difficult to establish, so we have elected not to attempt a closer fit at this time, but instead take the pragmatic
approach of generating what is most likely too many vacancy loops ( though still within a reasonable multiple ) in order to improve our
statistics.

\begin{figure}[h!t!b]
	        \begin {center}
	            \includegraphics [width=5.0in] {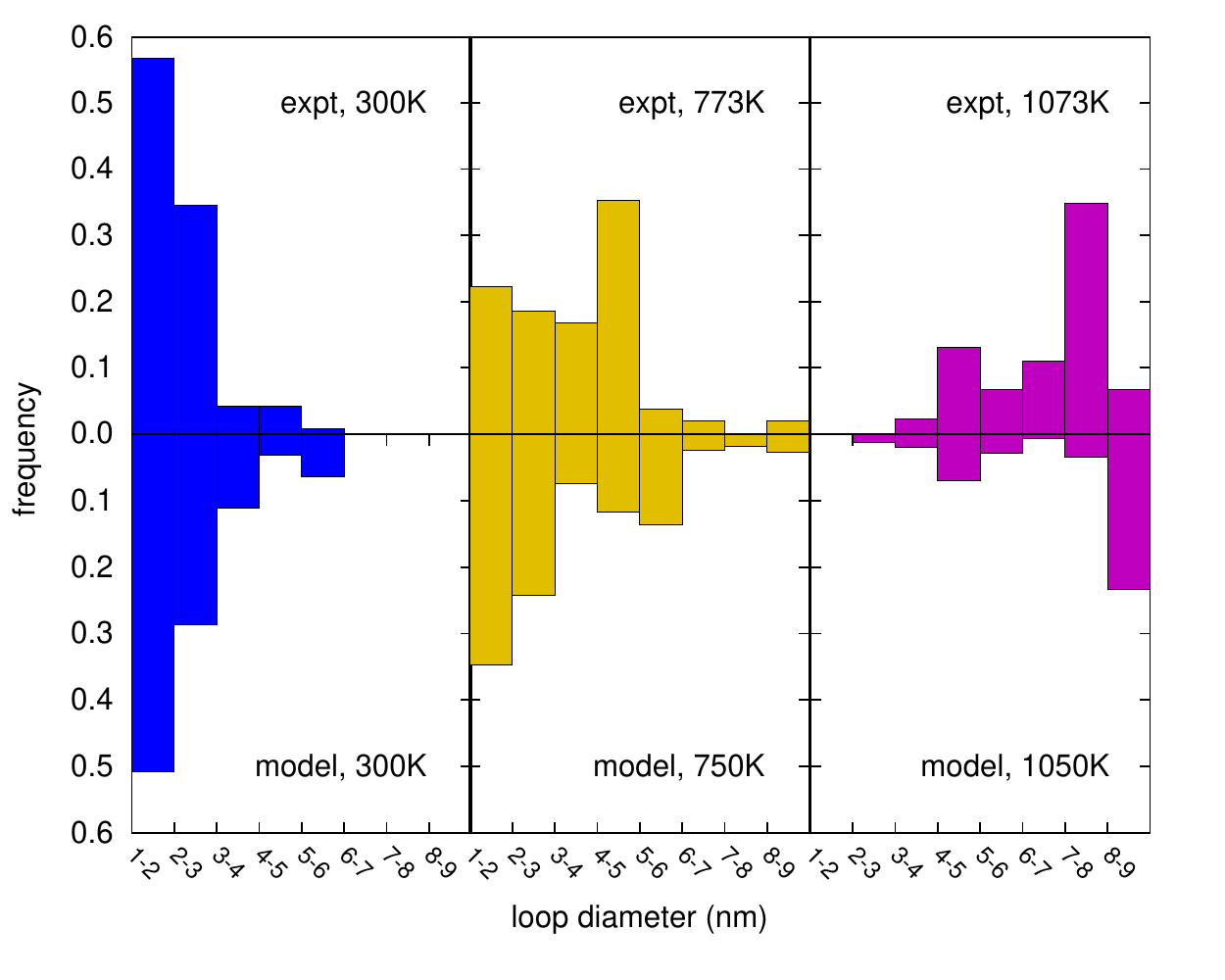}
	        \end{center}
	        \caption {
	        Histograms of the frequency of occurance of loops within a given diameter range, comparing the experimentally observed image size and the distribution remaining in the model at 1s.
	        Loops $>9$nm diameter were not binned in the simulations and only contributed to absolute numbers.
	        }
	        \label {loop_histogram}
	    \end {figure}

	    Histograms of the frequency of occurance of loops of different sizes are shown in figure \ref{loop_histogram}.
	    Here we see the clear correspondence between the experimental observations and the simulated results.
	    At low temperatures we see very few large loops remain to be observable, at intermediate temperatures there is a peak in the distribution for loop diameters around 5nm, and at high temperatures very few small loops are seen.

	We have therefore made a direct link between the defect size distributions generated in MD simulations and the defect size distributions seen in experiment, using object Kinetic Monte Carlo for the temperature dependent microstructural evolution, with no parameter tuning.

	\section{Discussion of results}

	    \subsection*{Comparison to resistivity recovery experiments}

	    Residual electrical resistivity recovery experiments performed on neutron-irradiated tungsten\cite{Thompson_pm1958,Thompson_pm1960}
show characteristic stages of recovery identified by isochronal annealing.
	    The first, below -170C, is attributed to the movement of essentially free interstitials.
	    Stage II, between -170C and 350C is a steady recovery attributed to the release of interstitials from traps in a wide range range of
energies from 0.25-1.7eV. This is supported by the increased stage II recovery in a cold-worked sample where a higher concentration of sinks
is assumed to be present.
	    Stage III is a rapid recovery from 350C to 450C due to monovacancy mobility becoming activated.

	    Our modelling shows that elastic trapping of loops provides activation barriers to long-range loop migration at the energy range of
stage II recovery.
	    The fact we do not see a dramatic high temperature effect due to vacancy mobility requires explanation:
	    In our randomly generated configurations about half the vacancies are found as monovacancies (see Appendix \ref{initialDistribution}).
	    The average number of Frenkel pairs per cascade is 200, so there are about 100 free monovacancies per cascade on average.
	    These will not be visible within the electron microscope unless they coalesce.
	    Our experiments see the average number of vacancies per loop increase from 60 vacancies per loop at room temperature to 774 vacancies per loop at 1073 K.
	    So even if all the free vacancies are drawn into loops ( a highly unlikely scenario ), it will make only a modest contribution to the
growth of the ``average'' loop.
	    In practice vacancies are more likely to diffuse away or be annihilated by interstitials than to be drawn into loops.

	    In a resistivity recovery experiment, the monovacancies \emph{are} detectable. If half of the vacancies are monovacancies, then this
part of the resistivity will be recovered when they start to move to sinks.

    \subsection*{Comparison with other transition metals}

	    Elastic interaction between defects must occur in all materials, but the energy scale will vary.
	    We can make some rough estimates for defects in other irradiated metals as follows.

	    The average isotropic elastic interaction energy for two arbitrarily oriented $\half\langle111\rangle$ loops is zero, but the trap
depth is given by the most negative value.
	    The minimum value occurs where the Burger's vectors of the two loops are not parallel ( or a pair of loops can move co-operatively ),
and the cylinders swept out by each loop just touch ( or the pair could interact ).

	    To simplify the material dependence we will proceed by assuming a Poisson's ratio of $1/3$ and that materials are elastically
isotropic.
	    Equation \ref{Isotropic_elastic_loop_loop} then has minima at approximately
	    \begin{eqnarray}
	        U_{12} &\approx& - 0.027 \mu a_0^3 \frac{N_1 N_2}{ \left(\sqrt{N_1}+\sqrt{N_2} \right)^3 }       \quad \quad \mbox{( opposite
character loops )}  \nonumber\\
	        U_{12} &\approx& - 0.061 \mu a_0^3 \frac{N_1 N_2}{ \left(\sqrt{N_1}+\sqrt{N_2} \right)^3 }       \quad \quad \mbox{( same
character loops )}.
	    \end{eqnarray}
	    where $N_1,N_2$ are the numbers of point defects in each loop.
	    This result suggests same-character (ie vacancy-vacancy or interstitial-interstitial) loops are more strongly bound together, but in
our modelling we found the probability of generating a configuration with two large same character loops was very small.
	    We will continue with this rough estimate assuming it is opposite character loops which trap each other.
	    Trap energies based on these approximations are given in table \ref{elasticTrapDepths}.

        \begin{table}
            \begin{center} \small
               \begin{tabular}{l|l|llll|l}
               element  &      $\mu a_0^3$ &   $N = 50$      & $N = 100$   & $N = 200$   & $N = 500$      &    $H_V^m$\cite{Derlet_PRB2007}
               \\
                        &      eV          &   eV                  &   eV              &   eV              &   eV  & eV\\
                    \hline
                    V         &        8.1   &  0.19	&0.27&	0.39&	0.61   &       0.62                  \\
                    Fe        &       12.1   &  0.29	&0.41&	0.58&	0.91   &       0.64                  \\
                    Mo        &       24.7   &  0.59	&0.83&	1.18&	1.86   &       1.28                  \\
                    Ta        &       15.5   &  0.37	&0.52&	0.74&	1.17   &       1.48                  \\
                    W         &       31.6   &  0.75	&1.07&	1.51&	2.38   &       1.78                  \\
	   			\end{tabular}
            \end{center}
            \caption{
            	Rough estimates for elastic trap depths for equal sized vacancy and interstitial loops in different materials.
            }
            \label{elasticTrapDepths}
        \end{table}

	    We can compare the energy scales in table \ref{elasticTrapDepths} with stage II resistivity recovery data.
	    In the case of iron there is only a small gap between the migration energy for self-interstitials and vacancies, being
0.27eV\cite{Takaki_RadEff1983} and 0.64eV\cite{Derlet_PRB2007} respectively. There is some evidence of recovery between 170K and 220K in
neutron-irradiated iron\cite{Matsui_JNuclM1988} absent in electron-irradiated iron\cite{Takaki_RadEff1983}, but this would be difficult to
ascribe conclusively to elastic detrapping of loops.
	    Neutron-irradiated molybdenum shows a clear stage II recovery\cite{Keys_prl1969} very similar to tungsten, which is consistent with
the prediction of deep elastic traps.
	    More compelling, however, is the observation that very high purity neutron irradiated tantalum appears to have very little stage II
recovery\cite{Burger_pl1966}, which is consistent with our prediction, as the shear modulus of tantalum is less than half that of
tungsten.

    \section{Conclusion}

		In this paper we have presented experimental observations of dislocation loops produced during the early stages of self-ion
irradiation of ultra-pure tungsten foil.
		Results are presented for a fluence of $10^{16}$ W$^+$ ions per square metre ( corresponding to a SRIM estimate of 0.01 dpa ), well
below the threshhold for cascade overlap.
		The loops are seen to be predominantly $\half\langle111\rangle$ Burger's vector vacancy type dislocation loops.
		At room temperature 2.99\% of cascades produce a visible loop, and the average loop diameter is 2.07nm. No very large loops ($>$8nm
diameter) are seen.
		At 1073K only 0.2\% of cascades produce a visible loop, but now the average diameter is 7.55nm, and almost no small loops ($<$3nm
diameter) are seen.
		With very few vacancy loops or mobile vacancy clusters produced per cascade, and at low fluence, it is not possible to ascribe the
change in loop diameter to a growth of individual loops, but rather it must be that small loops persist at low temperature, and large loops
persist at high temperature.

		In our experiments prismatic loops are generated at the surface of a foil, and with few loops containing an impurity atom at the time of generation.
		We have considered the elastic interaction energy between loops and vacancy clusters, and shown that it is quite large enough to trap loops to experimentally observable times.
		We do not claim that impurities are unimportant to microstructural evolution, rather we idealise our modelling for an experiment where elastic interactions are very important and impurity effects are occurring on a longer time-scale than loop loss.

		Object kinetic Monte Carlo simulations have been undertaken with mobile, interacting prismatic dislocation loops and vacancies handled
together.
All parameters used have been extracted from \emph{ab initio} or empirical potential calculations- none have been tuned to improve the fit with the
experiment described here.
		Our oKMC shows that large loops are quickly annihilated at low temperature, as they draw in small loops with opposing character.
		At high temperature the entropy increase due to the diffusion of the small loops balances this effect and then dominates, so that both
small and large loops start to move.
		Elastic loop trapping then occurs within a cascade where two or more loops can glide to an energetically favourable position.
		The great majority of loops are lost to the surface without finding such a minimum.
		The increased elastic interaction between large loops means that only large loops are sufficiently deeply trapped to persist to long
times at high temperature.

		These results were found by evolving thousands of randomly generated starting configurations out to several seconds simulated time.
		This large number is essential to see the effects we have here- only a fraction of a percent of cascades produce any visible defects,
and we must gather statistics over the unlikely cases where large loops are randomly generated.
		For this reason it was necessary to use rules for the generation of cascades, as it would be prohibitively expensive to use molecular
dynamics to directly generate starting states.
		The size and distribution of defects produced in a cascade is critical for determining the elastic interaction between objects, and so evolution is very sensitive to assumptions made about cascade structure.
		We have tried to reproduce as nearly as possible generic observed features seen in MD, but improving the
description of cascade morphology is clearly an area for profitable future work.

    \section*{Acknowledgements}

    	The authors would like to thank Mike Jenkins and Mark Gilbert for helpful discussions.
		This work was part-funded by the RCUK Energy Programme [grant number EP/I501045] and by the EPSRC via a programme grant EP/G050031 and by the European Union's Horizon 2020 research and innovation programme.
		To obtain further information on the data and models underlying this paper please contact PublicationsManager@ccfe.ac.uk.
		The views and opinions expressed herein do not necessarily reflect those of the European Commission.

		The in-situ irradiation experiments were accomplished at the Electron Microscopy Center for Materials Research at Argonne National Laboratory, a US Department of Energy Office of Science Laboratory operated under Contract No.DE-AC02-06CH11357 by U. Chicago Argonne, LLC.
		We thank Pete Baldo for his help with the irradiations.
		Many thanks to the China Scholarship Council studentship funding (XY) and to the EPSRC for support for this research, via the program grant `Materials for Fusion and Fission Power', EP/H018921/1.

\bibliographystyle{unsrt}

\appendix

	\section{Initial distribution of objects}
			\label{initialDistribution}
		Interstitial loops are initially distributed with position and size uncorrelated
			\begin{equation}
				P_{\mbox{int}}(N,R) \sim \left( \frac{1}{N^{s_I}} \right) \exp{ \left[ - \frac{ (R - \sigma)^2 }{2 \sigma^2} \right] },
			\end{equation}
		with the spatial extent of the cascade defined by $\sigma$.
		We take $\sigma = 5$nm.
		Vacancy loops and clusters are distributed with larger loops in the centre and smaller clusters outside using the formula
			\begin{equation}
				p_{\mbox{vac}}(N,R) \sim  \exp{ \left[ - \frac{1}{2} \left( \frac{ R }{ \sigma} \right)^{s_V} N \right] } \exp{ \left[ -
\frac{ R^2 }{2 \sigma^2} \right] }.
			\end{equation}
		A vacancy object with $N\le 13$ is set to be a cluster, and $N>13$ set to be a vacancy loop.
		A configuration of $M$ loops and clusters is randomly drawn from these distributions, and the numbers of vacancies and interstitials
is compared.
		Additional vacancy or interstitial objects are then drawn, whichever is in deficit, using the same distributions.
		If an equal number of vacancies and interstitials is generated, the drawing ends, but if a surplus of the opposite kind is generated,
the configuration is rejected.
		To better match the sparseness of the vacancy clusters seen in ref\cite{Sand_EPL2013}, vacancy clusters are split and redistributed
over twice their minimum volume.
		Finally the configuration is then tested for overlap between objects, and the loops and clusters combined appropriately.

		The initial number of loops drawn $M$ and the power indices $s_I$ and $s_V$ are varied to reproduce as closely as possible the
distribution of loops found in MD simulations by Sand et al.\cite{Sand_EPL2013}.
		We choose to make the distribution of sizes of vacancy and interstitial objects as closely matched as possible, for want of better
information.
	       \begin{figure}[h!t!b]
	           \begin {center}
	               \includegraphics [width=5.0in] {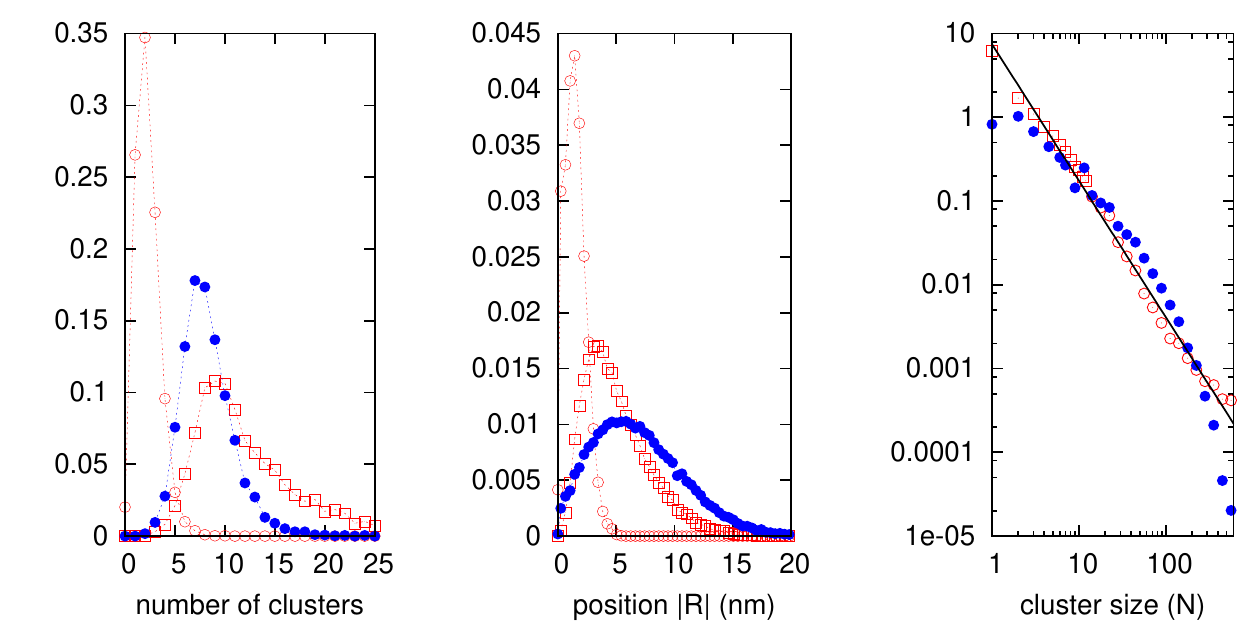}
	           \end{center}
	           \caption {
	           Statistics for configurations of loops generated using the procedure in Appendix \ref{initialDistribution}.
	           Open circles: vacancy loops. Open squares: vacancy clusters. Closed circles: interstitial loops.
	           Left: the number of each type of object generated in one cascade.
	           Centre: the distance from the centre of the cascade.
	           Right: the number of defects in each object. The solid black line is the power-law distribution of interstitial loops from
ref\cite{Sand_EPL2013}, $p(N)=7.45 N^{-1.63}$.
	           	}
	           \label {cluster_config}
	       \end {figure}
		Histograms of the distributions of the objects are shown in figure \ref {cluster_config}.
		We find a good fit where $s_I=1.63, s_V=1.37, M= 22$.

    \section{Parameterization of OKMC}
    \label{parameterization}
    	\subsection*{Geometric properties of objects}
        \begin{table}
            \begin{center} \small
                \begin{tabular}{l|ll}
                    &   Vacancy clusters    & $1/2 <111>$ Loops  \\
                    \hline
                       $\rho_N$ & $a_0 \left( \frac{ 3 N }{ 8 \pi } \right)^{1/3}$	& $a_0 \sqrt{ \frac{ N }{ \pi\sqrt{3} } }$ \\
	   				   $\Delta \rho$ & $0.5 a_0$	& $a_0$ \\
       				   $\Delta V (N=1)$     &   $-0.37 \Omega $      \\
	   				   $\Delta V (N>1)$     &   $-0.531 N^{2/3} \Omega $
	   			\end{tabular}
            \end{center}
            \caption{
            	Fundamental geometric properties of objects: the radius of an object containing $N$ point defects $\rho_N$, the capture radius
            $\Delta \rho$ and the relaxation volume $\Delta V$.
            	$\Omega = a_0^3/2$ is the volume per atom.
            }
            \label{GeometryOKMC}
        \end{table}

        Collisions between objects are determined based on geometric idealisations of voids as spheres and loops as tori.
        Spheres are modelled with a radius $\rho + \Delta \rho$, where $\Delta \rho$ is an additional capture radius,
        tori have a major radius $\rho$ and minor radius $\Delta \rho$.
        The values for these as a function of number of defects contained is given in table \ref{GeometryOKMC}.

        Voids interact if their spheres intersect, and a void interacts with a loop if the void's sphere intersects the loop's torus.

        A pair of loops are deemed to collide if the sphere circumscribing one loop intersects the torus of the other, \emph{and vice versa}.
        This is not quite the same as a criterion where two tori intersect, but it is preferred as a collision is also claimed when two
        parallel loops stack together with their centres only slightly displaced.

    	\subsection*{Formation energy and migration energy of objects}

    Formation free energies (ie including vibrational entropies but excluding configurational entropies) of the defects have been computed for
    the objects using lattice statics with an empirical potential\cite{mason_2014}.
    The temperature variation was assessed as a polynomial fit to calculations in the range T=0 to T=1800K, with a supercell of 24x24x24 unit
    cells ( 27.6k atoms ), and the defect size variation fitted to established formulae \cite{Dudarev_prl2008}.
    The vacancy cluster energies (N=3-13) were fitted to the lowest energy of hundreds of structures, and the energy of vacancy voids to
    $<110>$ facetted rhombic dodecahedra, known to be the lowest energy void structures\cite{Rasch_PMA1980}.
    The void energy expression contains a small constant term to match with clusters at N=13.
    Formation free energies are listed in table \ref{ParameterizationOKMC}, together with parameterizations for rates of events.

        \begin{table}
            \begin{center} \small
                \begin{tabular}{l|l}
                        &   Vacancy clusters / voids   \\
                    \hline                  \\  [0.5ex]
   				   $G^f (N=1)$     &   3.743 - 1.781$k_B T$ + 8.279($k_B T$)$^2$                                       \\
   				   $G^f (N=2)$        &   7.335 - 2.810$k_B T$ + 14.063($k_B T$)$^2$                                      \\
   				   $G^f (3\le N \le 13)$&  (2.479 - 0.992$k_B T$ + 3.8252($k_B T$)$^2$)N     + ( 3.810 - 0.940$k_B T$ + 4.059($k_B T$)$^2$)
   \\
   				   $G^f (N \ge 13)$    &   ( 6.543 - 0.664$k_B T$ + 10.74($k_B T$)$^2$ ) N$^{2/3}$ + ( -0.136 - 10.16$k_B T$ - 5.59($k_B
   T$)$^2$)                        \\ [0.5ex]
   				   $\Gamma_{\mbox{3D}} (N=1)$  & $8 \times  4 \nu_0 \exp\left[ -H^m_V/k_B T \right]$
   \\
   				   $\Gamma_{\mbox{3D}} (1<N<13)$  &  $\frac{\rho_N^2}{\rho_1^2} \times \nu_0 \exp\left[ -H^m_V/k_B T \right]$              \\
   				   $\Gamma_{\mbox{3D}} (N\ge13)$  &  $\frac{\rho_{13}^2}{\rho_1^2} \times \nu_0 \exp\left[ -H^m_V/k_B T \right]$
   \\  [0.5ex]
   				      				  \\ [0.5ex]
   				   \hline
   				   \multicolumn{2}{c}{}\\
   				       &   Crowdions          \\
   				   \hline                      \\  [0.5ex]
   				    G$^f$           &  ( 8.300 - 0.105$k_B T$ )    \\ [0.5ex]
   				    $\Gamma_{\mbox{1D},\parallel}$    &   $2 \times \nu_0 \exp\left[ -H^m_I/k_B T \right]$                        \\ [0.5ex]
   				    $\Gamma_{\mbox{rotation}}$    &   $3 \times \nu_0 \exp\left[ -H^r_I/k_B T \right]$                        \\ [0.5ex]
   				   \hline
   				   \multicolumn{2}{c}{}\\
   				       &   Vacancy loops       \\
   				   \hline                      \\  [0.5ex]
   				    G$^f$   &   ( 7.053 - 21.85$k_B T$ )$\sqrt{N} \log{\sqrt{N}}$ + ( 3.565 + 6.073$k_B T$ )$\sqrt{N}$              \\
   				    $\Gamma_{\mbox{split } N \rightarrow (N-1) + 1}$    &   $\frac{\rho_N}{\rho_1} \times \nu_0  \exp\left[ -H^m_V/k_B T
   \right]$                        \\ [0.5ex]
   				   \hline
   				   \multicolumn{2}{c}{}\\
   				       &   Interstitial loops          \\
   				   \hline                      \\  [0.5ex]
   				    G$^f$           &   ( 6.470 + 4.088$k_B T$ - 0.0168($k_B T$)$^2$ )$\sqrt{N} \log{\sqrt{N}}$ + ( 8.300 - 0.105$k_B T$
   )$\sqrt{N}$    \\
   				    $\Gamma_{\mbox{rotation}} (2\le N \le3)$    &   $3 \times \nu_0 \exp\left[ -H^m_V/k_B T \right]$                        \\
   [0.5ex]
   				   \multicolumn{2}{c}{}\\
   				       &   All $1/2 <111>$ loops          \\
   				   \hline                      \\  [0.5ex]
   				    $\Gamma_{\mbox{1D}\parallel}$    &   $2 \times \frac{\displaystyle \rho_1}{\displaystyle \rho_N} \nu_0 \exp\left[
   -H^m_L/k_B T \right]$                        \\
   				    $\Gamma_{\mbox{2D}\perp}$    &   $\frac{\rho_N}{\rho_1} \times \nu_0  \exp\left[ -H^m_V/k_B T \right]$
   \\ [0.5ex]
                \end{tabular}
            \end{center}
            \caption{
            	Parameterization of the free energy of formation and migration rates for the three classes of objects.
            	Vacancy clusters are permitted to undergo Brownian motion, with a steplength of $a\sqrt{3}/4 N$, in eight $\langle111\rangle$
            hop directions.
            	The total rate for any such hop is shown.
            	It is worth noting the crowdion formation energy is somewhat lower than the accepted DFT value ( 9.55-10.53 eV
            )\cite{Derlet_PRB2007,Marinica_JPCM2013} in a box of 128+1 atoms.
            	We keep this number to provide consistency with larger interstitial cluster energies.
            	Interstitial rotation is for the principal axis changing from $\half[111]$ through a $\langle110\rangle$ direction to another
            $\half\langle111\rangle$.
            	Loop Brownian motion and splitting is discussed in the text.
            }
            \label{ParameterizationOKMC}
        \end{table}
    	The following thermal activation barriers are used:
    	for monovacancy migration $H^m_V =	1.78$eV \cite{Muzyk_PRB2011},
    	for crowdion migration $H^m_I =	0.01$\cite{Derlet_PRB2007,Amino_PML2011},
    	for crowdion rotation $H^R_I = 0.38$\cite{Derlet_PRB2007}, and for loop migration $H^m_L = 0.13$eV.
    	This last value was computed as an Arrhenius fit to MD simulations of the diffusion of an isolated 55 vacancy loop, using the method
    described in ref\cite{Derlet_PRB2011}.
    	The rate prefactor $\nu_0$ is taken to be the Debye frequency, $\nu_0 = 6.45 \times 10^{12} s^{-1}$\cite{Becquart_JNM2010}.

\end{document}